\documentclass[aps,pre,groupedaddress,showkeys,showpacs,amsmath]{revtex4-1}
\usepackage{amssymb}
\usepackage{amsmath}
\usepackage{graphicx}
\usepackage{amsbsy}
\usepackage{xr}
\usepackage{float}
\usepackage{bm}
\usepackage{color}
\usepackage{gensymb}
\usepackage{caption}
\usepackage{subcaption}
\usepackage[space]{grffile}
\definecolor{cream}{RGB}{222,217,201}
\usepackage{siunitx}
\DeclareSIUnit[number-unit-product = {\,}]
\cal{cal}
\DeclareSIUnit\kcal{\kilo\cal}
\DeclareSIUnit\kcal{\kilo\joule\per\mole}
\DeclareSIUnit\molar{\mole\per\cubic\deci\metre}
\DeclareSIUnit\Molar{\textsc{m}}
\usepackage{multirow}

\begin{document}

\title{Chain stiffness bridges conventional polymer and bio-molecular phases}

\author{Tatjana \v{S}krbi\'{c}} 
\email{tskrbic@uoregon.edu}
\affiliation{Department of Physics and Institute for Theoretical Science, 1274 University of Oregon, Eugene, OR 97403-1274, USA}
\affiliation{Dipartimento di Scienze Molecolari e Nanosistemi, 
Universit\`{a} Ca' Foscari di Venezia
Campus Scientifico, Edificio Alfa,
via Torino 155, 30170 Venezia Mestre, Italy}
\author{Jayanth R. Banavar}
\email{banavar@uoregon.edu}
\affiliation{Department of Physics and Institute for Theoretical Science, 1274 University of Oregon, 
Eugene, OR 97403-1274, USA}

\author{Achille Giacometti} 
\email{achille.giacometti@unive.it}
\affiliation{Dipartimento di Scienze Molecolari e Nanosistemi, 
Universit\`{a} Ca' Foscari di Venezia
Campus Scientifico, Edificio Alfa,
via Torino 155, 30170 Venezia Mestre, Italy}
\affiliation{European Centre for Living Technology (ECLT)
Ca' Bottacin, 3911 Dorsoduro Calle Crosera, 
30123 Venice, Italy}

\date{\today}

\begin{abstract}
Chain molecules play important roles in industry and in living cells. Our focus here is on distinct ways of modeling the stiffness inherent in a chain molecule. We consider three types of stiffnesses -- one yielding an energy penalty for local bends (energetic stiffness) and the other two forbidding certain classes of chain conformations (entropic stiffness). Using detailed Wang-Landau microcanonical Monte Carlo simulations, we study the interplay between the nature of the stiffness and the ground state conformation of a self-attracting chain. We find a wide range of ground state conformations including a coil, a globule, a toroid, rods, helices, zig-zag strands resembling $\beta$-sheets, as well as knotted conformations allowing us to bridge conventional polymer phases and biomolecular phases. An analytical mapping is derived between the persistence lengths stemming from energetic and entropic stiffness. Our study shows unambiguously that different stiffness play different physical roles and have very distinct effects on the nature of the ground state of the conformation of a chain, even if they lead to identical persistence lengths.
\end{abstract}
\maketitle

\section{Introduction}
\label{sec:introduction}
Linear chain molecules adopt a wide range of ground state conformations, which play important roles in industry and in life. Proteins are relatively short chain molecules made up of 20 kinds of amino acids that curl up under self-attraction into conformations comprised of assemblies of tightly wound $\alpha$-helices and almost planar $\beta$-sheets \cite{Lesk10,Finkelstein16}. Because a protein folds reproducibly and rapidly into its native state conformation independently of the specific sequence of the amino acids, the folded state is generically postulated to be a free energy minimum or a ground state. For proteins, the folded form determines function. Intriguingly, in spite of the dynamical challenge of folding, a fraction of the folded conformations of proteins are now known to have knots \cite{Micheletti11,Chwastyk15,Alexander17}. Interestingly, the structures in conventional polymer physics are distinct from those of biomolecules and include a coil (loosely analogous to an unfolded protein structure), an unstructured compact globule, rods assembled approximately parallel to each other, and a toroid. From a theoretical and modeling point of view, it is desirable to capture the essential attributes of a chain molecule in the simplest manner, while yet retaining the rich generic behavior of the diverse systems. Here we show unambiguously, through extensive computer simulations, that different kinds of stiffnesses play different physical roles, they are not interchangeable, and they have very distinct effects on the nature of the zero temperature (ground state) conformation of a chain. Furthermore, our studies shed light on the existence of knotted ground states. Our work is not only important in polymer physics but is also directly relevant to understanding proteins, the workhorse molecules of life, as well as other biomolecules.

The simplest standard model of a linear chain molecule \cite{Taylor09b,Bhattacharjee13} is a railway train arrangement of tethered hard spheres of diameter $\sigma$ with consecutive spheres tangent to each other and barely touching. This automatically leads to an effective stiffness (an effect that will be accentuated later in our study) -- conformations with a local bending angle greater than 120$\degree$ are forbidden. (It is important to note that a fully flexible ball and stick model does not have such an effective stiffness.) We will refer to this as \textit{entropic stiffness} as its origin is in the reduction of the available conformational space. One postulates an attractive interaction between non-consecutive spheres -- in the simplest case, there is a unit energy gain if two sphere centers are within a threshold range defined to be $R_c$. The ground states of this model include an unfolded coil and a compact unstructured globule (or a structured crystal). A flaw of this simple model is the incompatibility between the spherical symmetry of the constituent spheres and the uniaxial local symmetry of a railway train. Our goal is to overtly incorporate the uniaxial symmetry inherent in a chain through an effective (entropic) stiffness. 

We consider three ways of introducing stiffness that have been presented in the literature, each of which respects the uniaxial symmetry of a chain. The first is a bending energy penalty yielding the so-called worm-like chain model. A significant fraction of synthetic polymers are known to exhibit some rigidity to bending that is usually referred to as `stiffness' \cite{Flory69,Grosberg94,Rubinstein03}. This is captured through an energetic penalty associated with each bending angle of a chain, unlike its flexible counterpart that can be bent at no energy cost. One invokes an elastic bending energy, that is zero for a straight chain (the local bending angle is 0$\degree$ in this case) and increases quadratically with the local bending angle $\theta$ (Figure \ref{fig:fig1a}). A more realistic description that avoids the unrealistic growth of the elastic energy at large bending angle, is to use a $1-\cos\theta$ dependence that coincides with the quadratic cost for small $\theta$ but flattens out (recall that a $\theta$ angle greater than 120$\degree$ is forbidden in any case). This is the basis of the so-called Kratky-Porod model that has been particularly useful in theoretical work both at the discrete and continuum levels \cite{Rubinstein03,Bhattacharjee13} and takes the more common name of the worm-like-chain model (WLC see Figure \ref{fig:fig1a}). This stiffness and the other two types we consider are characterized by a persistence length $L_p$, so that the polymer is locally stiff but becomes flexible at length scales larger than $L_p$, hence the name semi-flexible polymers. Besides synthetic polymers, several biopolymers belong to the category of semi-flexible polymers. This includes, for instance double stranded DNA \cite{Vologodskii15} for which the persistence length is known to be 150bp (equivalent to $\approx 50$nm). The WLC model fully accounts for the experimental findings in DNA condensation that can be easily rationalized in terms of competition between the hydrophobic effect, promoting compaction, and the bending rigidity, tending to oppose it \cite{Hoang14,Hoang15}. We will refer to this more conventional stiffness as \textit{energetic}, in order to distinguish it from the entropic one.

Proteins are polypeptide chains with no significant intrinsic energetic bending rigidity \cite{Lesk10,Finkelstein16}, although the WLC model is still useful within this framework to describe, for instance, protein self-assembly in amyloids (see \cite{Ranganathan16} for a recent example). At the level of single chain, however, the WLC model cannot account for the formation of characteristic secondary structures found in real proteins, such as $\alpha$-helices, $\beta$-sheets, and combinations of the two. Yet, proteins do have restrictions to their bending stemming from the presence of side chains that typically stick out from the protein backbone along a direction roughly in the negative normal direction \cite{Finkelstein16}, but the origin and physical effect of this stiffness is different from that described earlier. This is the motivation for an entropic stiffness, that we study here, which can be modeled by attaching hard side spheres (diameter given by $\sigma_{sc}$, tangentially to all but the first and last main chain spheres, in the negative normal direction -- the polymer-with-side-chains model (PSC Figure \ref{fig:fig1b}). These  side spheres are not allowed to overlap with the main chain spheres or with each other and serve to thin the available phase space of conformations.  

Another kind of entropic stiffness arises when the constituent consecutive spheres along a chain are not barely touching (or tangent to each other) as in a standard model but are allowed to overlap (the OP model standing for overlapping polymer (Figure \ref{fig:fig1c}). This reduces the bond length $b$ and brings the centers of consecutive spheres closer to each other than the sphere diameter $\sigma$ thus enabling more pairs of spheres to avail of the attractive interaction for a given fixed range of attraction $R_c$. Equally importantly, it broadens the scope of the constraint on the local bending angles of a chain conformation. There is no energy cost for local bending angles smaller than a threshold angle (that can now be less than 120$\degree$) but sharper bends are forbidden. This again reduces the phase space that a chain can explore and hence the terminology entropic stiffness \cite{ Skrbic16a,Skrbic16b}.

A word of caution is in order at this point. The stiffness of a real synthetic chains is determined by its structure as well as by hindrance to bond rotation. Even in proteins, the planar configurations of the atomic groups, due to specific bond hybridization, allows the introduction of virtual bonds connecting the neighboring carbon atoms \cite{Flory69,Volkenstein63}. The polymer literature abounds with references \cite{Flory69,Grosberg94,Rubinstein03,Volkenstein63} to the physical origins of chain stiffness and a recognition that the concepts of entropic and energetic stiffnesses are intertwined.

The models considered here are classical polymer physics models and follow the expectations established by Flory, DeGennes, and numerous others in the literature \cite{Flory69,Grosberg94,Rubinstein03}. We do wish to stress that our focus in this paper is on the behavior of a single chain molecule and not on a concentrated system. Furthermore, we restrict ourselves to the study relatively short chain molecules because of their relevance to proteins. 

While the building blocks of protein structures (helices and almost planar sheets) are not observed in the WLC model, a signature of the presence of helical phases is seen in the OP model \cite{Clementi98,Magee06} and both helices and $\beta$-like structures are observed in the PSC model (Figure \ref{fig:fig2}). The notion of entropic stiffness originates from the modeling of proteins using the tube or thick-chain model \cite{Maritan00,Hoang04,Banavar07}.  Recent work on combining the effects of the OP and the PSC stiffnesses (overlapping-polymer-with-side-chains OPSC model Figure \ref{fig:fig1d}) has revealed a remarkably rich ground state phase diagram akin to biomolecular phases. In particular, a novel phase denoted as the ``elixir phase'' was found which has ground state conformations assembled from helices and sheets as a result of a further reduction of the symmetry \cite{Skrbic19b}. Indeed, while respecting the original uniaxial symmetry of the chain (as in the OP case), this model has biaxial symmetry due to the addition of the side chain. As a result, an additional phase (the elixir phase) emerges in a way akin to what happens in liquid crystals \cite{Chaikin00}. The remarkable properties of this phase suggest that globular protein native state structures might reside in this phase thereby possessing their many amazing common characteristics \cite{Skrbic19a,Skrbic19b}. The PSC model, on the other hand, does \textit{not} respect the uniaxial symmetry of the chain, and hence does not share this property, as we shall see. Our study here is the first, to our knowledge, that addresses all three kinds of stiffnesses, and their role in determining the ground state conformations of a chain molecule including the presence of knotted conformations. Figure \ref{fig:fig1} underscores the differences between the three types of stiffnesses encapsulated in the models.

The energetic stiffness arises from an energy penalty for bending and contributes to the energy of a conformation. The degree to which a given energy is prohibitive depends on the temperature. At very high temperature, the energy does not matter as much. In contrast, at very low temperatures, the ground state is an allowed conformation with the lowest energy. The overlap leads to an effective potential which is zero when the bending is not too stringent and prohibitively expensive when the bending is tighter than a given threshold. This 0-infinity potential is temperature independent and merely thins the conformational space and hence is entropic in origin. Even more clear is the entropic stiffness resulting from the presence of non-overlapping hard side spheres. Here again the potential is infinity when overlap occurs and is zero otherwise and in a temperature independent manner.

The plan of the paper is as follows. Section \ref{sec:models} presents the details of the different stiffness models, and Section \ref{sec:simulations} the methods used for numerical simulations and data analysis. Section \ref{sec:results} presents the numerical results, and Section \ref{sec:different} an analytical mapping between the persistence lengths of different models. Section \ref{sec:conclusions} finally draw some conclusions.

  \section{The models}
  \label{sec:models}
\begin{figure}[htpb]
  \centering
  \captionsetup{justification=raggedright,width=\linewidth}
  \begin{subfigure}{8cm}
   \includegraphics[trim = 30mm 40mm 10mm 10mm, clip,width=\linewidth]{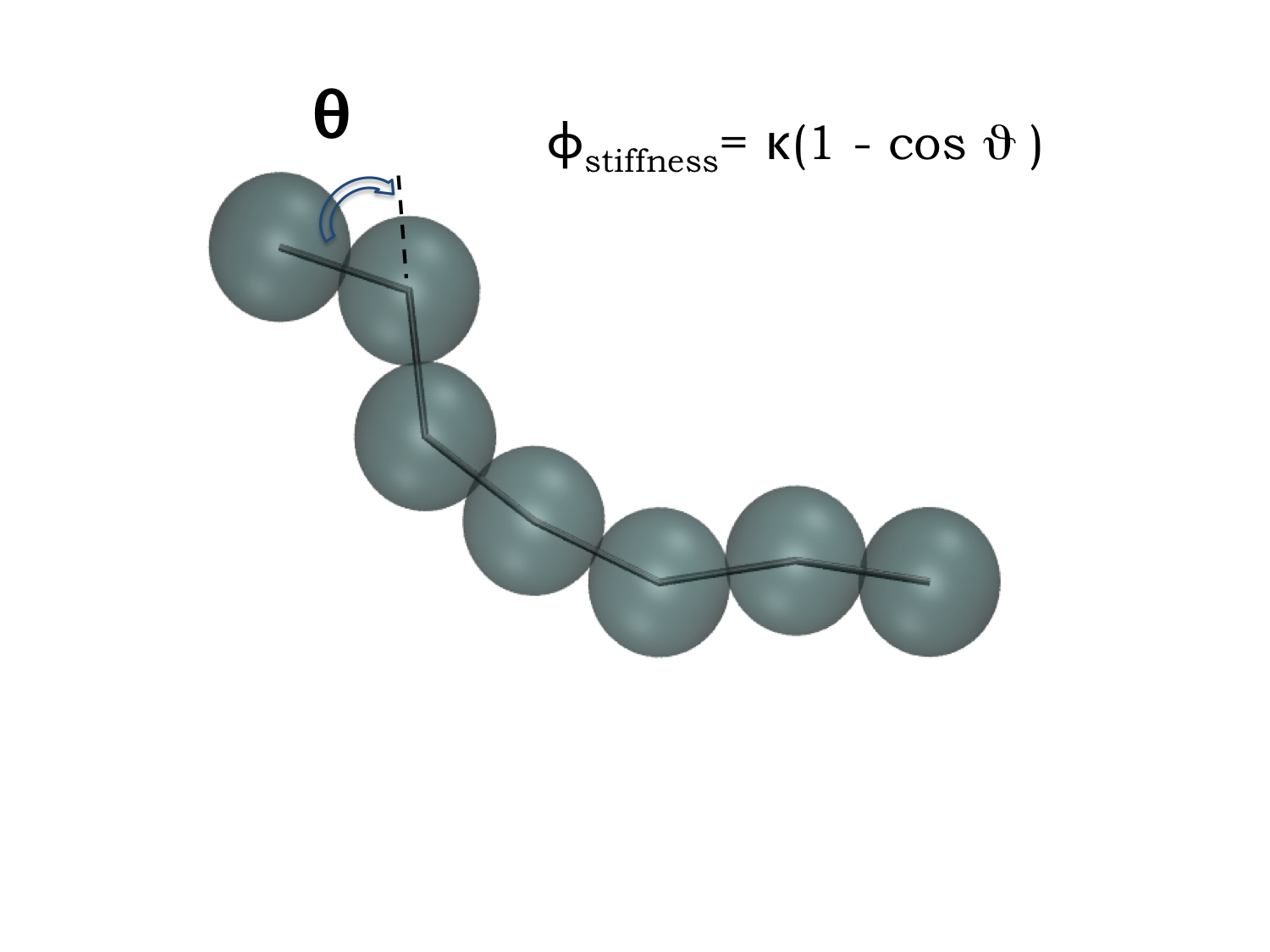}
    \caption{}\label{fig:fig1a}
  \end{subfigure}
  \begin{subfigure}{6cm}
    \includegraphics[width=\linewidth]{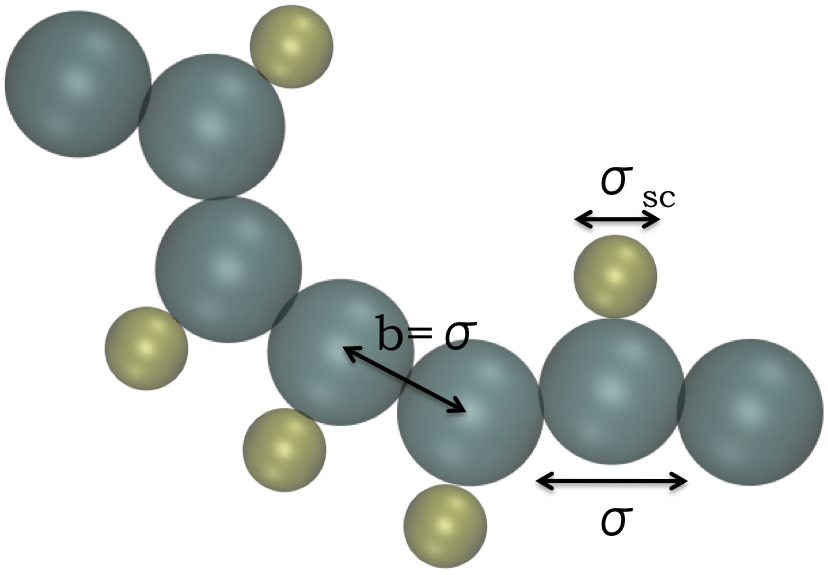}
    \caption{}\label{fig:fig1b}
  \end{subfigure}\\
  \begin{subfigure}{6cm}
    \includegraphics[width=0.8\linewidth]{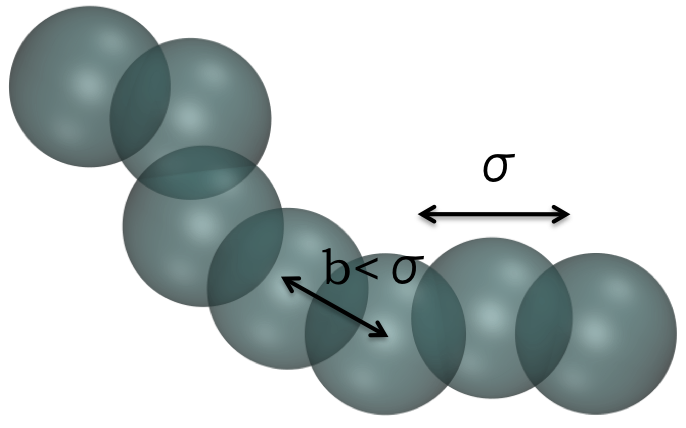}
    \caption{}\label{fig:fig1c}
  \end{subfigure}
  \begin{subfigure}{6cm}
   \includegraphics[width=0.75\linewidth]{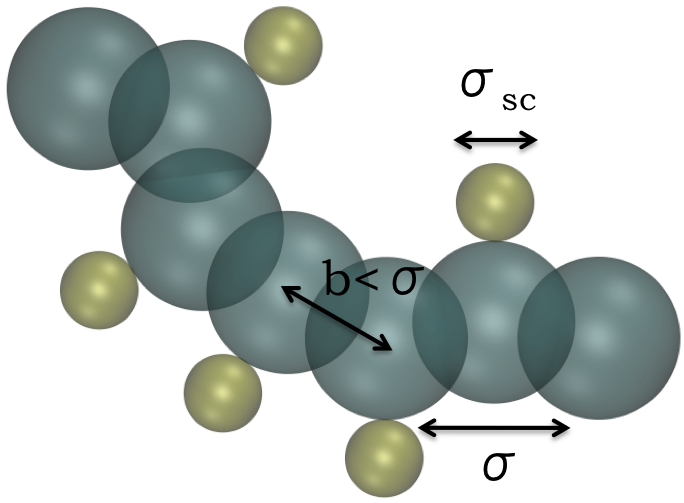}
    \caption{}\label{fig:fig1d}
  \end{subfigure}

  \caption{Different models for capturing the effects of chain stiffness. $b$ is the bond length, the distance between the centers of adjacent spheres along the chain. $\sigma$ and $\sigma_{sc}$ are the diameters of the main chain and side spheres. $\kappa$ is the energetic stiffness parameter and $\epsilon$ is a characteristic energy scale, equal to the strength of the pairwise attractive interaction between main-chain spheres.  Here we set $\epsilon = 1$ without loss of generality. (a) The  worm-like-chain (WLC) model $b/\sigma=1$, $\sigma_{sc}/\sigma=0$, $\kappa/\epsilon>0$; (b) The PSC model $b/\sigma=1$, $\sigma_{sc}/\sigma>0$, $\kappa/\epsilon=0$; (c) The OP model $b/\sigma<1$, $\sigma_{sc}/\sigma=0$, $\kappa/\epsilon=0$; (d) The OPSC model $b/\sigma<1$, $\sigma_{sc}/\sigma>0$, $\kappa/\epsilon=0$.
  \label{fig:fig1}}
\end{figure}
Consider a chain formed by $N$ spherical beads located at positions $\mathbf{R}_{1},\ldots,\mathbf{R}_{N}$, and let $\mathbf{r}_{ij}=\mathbf{R}_j-\mathbf{R}_i$ be the vector distance between two beads. Let $r_{i,i-1}=\vert \mathbf{r}_{i,i-1} \vert =b$ be the distance between two consecutive backbone beads, and $\sigma$ be the diameter of the backbone bead. In the simplest case \cite{Taylor09b} $b=\sigma$ and adjacent beads are tangent to one another.

Non-consecutive spherical beads are subject to a square-well potential:
\begin{equation}
\phi\left(r\right)=\begin{cases}
+\infty\,,\quad \,r < \sigma &\\
-\epsilon,\quad\, \sigma < r < R_c&\\
0, \qquad r > R_c&  .
\end{cases}
\label{supp1:eq1}
\end{equation}
Here $\epsilon$ is taken to be 1 without loss of generality and the reduced temperature $T^{*} = k_BT/\epsilon$ quantifies the effective strength of the interactions.

In order to define the bending rigidity, let $\widehat{\mathbf{T}}_i$ be the local tangent along the chain axis at the $i$-th bead, so that the bending angle $\theta_i$ between the
$i$-th and the $(i+1)$-th bead is given by
\begin{eqnarray}
 \label{supp1:eq3}
 \widehat{\mathbf{T}}_{i+1} \cdot \widehat{\mathbf{T}}_{i} &=& \cos \theta_i
\end{eqnarray}
The WLC model then considers a confining potential of the form
\begin{eqnarray}
 \label{supp1:eq4}
 \phi_{\text{WLC}}\left(\theta\right) &=& \kappa \left(1-\cos \theta \right)
\end{eqnarray}
where $\kappa$ is the bending energy. The particular choice of the $\theta$ dependence in Eq.\ref{supp1:eq4} ensures elastic behavior at small $\theta$
as well as a flattening of this energy at larger angles. The total stiffness energy is then
\begin{eqnarray}
 \label{supp1:eq5}
	E_{\text{stiffness}} &=& \kappa \sum_{j=1}^{N-1} \left(1-\cos \theta_j\right)
\end{eqnarray}
As this kind of stiffness originates from an explicit potential term, we denote it as 'energetic stiffness'. Another kind of stiffness stems for forbidding certain classes of chain conformations. This effectively translates into a reduction of the conformational entropy of the chain, and hence will be denoted as 'entropic stiffness' in the following. We consider here three different ways of achieving this. In the first case, the polymer-with-side chains (PSC) model, side spheres of diameter $\sigma_{sc}$ are located at positions
\begin{eqnarray}
 \label{supp1:eq2}
\mathbf{r}^{(SC)}_i &=& \mathbf{R}_i - \widehat{\mathbf{N}}_i \left(\sigma + \sigma_{sc} \right)/2 .
\end{eqnarray}
where $\widehat{\mathbf{N}}_i$ is the normal unit vector at the $i-$th bead within the discrete Frenet frame (see below). In the second case, the overlapping-polymer (OP) model, adjacent beads are allowed to a partial interpenetration so that $\sigma/2< b < \sigma$ . In the last case, the overlapping-polymer-with-side-chains (OPSC) model, the two above effects are combined.

In the PSC and OPSC models, we consider sidechain spheres of hardcore diameter $\sigma_{sc}$ in the plane perpendicular to the chain axis.
To this aim, it proves convenient to introduce the discrete counterpart of the Frenet frame.
For each of the non-terminal backbone beads, one defines a tangent and a normal vector as
\begin{eqnarray}
\label{supp2:eq1}
\widehat{\mathbf{T}}_i & = & \frac{\mathbf{R}_{i+1} - \mathbf{R}_{i-1}}
{\left \vert\mathbf{R}_{i+1} - \mathbf{R}_{i-1} \right \vert} \\
\widehat{\mathbf{N}}_i & = & \frac{\mathbf{R}_{i+1}-2\mathbf{R}_i+\mathbf{R}_{i-1}}
       {\left \vert \mathbf{R}_{i+1}-2\mathbf{R}_i+\mathbf{R}_{i-1} \right \vert}  ,
\end{eqnarray} 
where $i=2,\ldots,N-1$. Consequently, one can also define a binormal vector as:
\begin{equation}
\label{supp2:eq2}
\widehat{\mathbf{B}}_i = \widehat{\mathbf{T}}_i \times \widehat{\mathbf{N}}_i .
\end{equation}
Note that $\{\widehat{\mathbf{B}}_i,\widehat{\mathbf{T}}_i,\widehat{\mathbf{N}}_i\}$, ($i=2,\ldots,N-1$) represent the discretized version of the Frenet-Serret local coordinate frame routinely used in polymer theory.

In summary, we have four different models:
\begin{itemize}
\item[WLC] Worm-Like-Chain model. Here $b/\sigma=1$, $\sigma_{sc}/\sigma=0$, $\kappa/\epsilon>0$; see Figure \ref{fig:fig1a}.
\item[PSC] Polymer-With-Side Chains model. Here $b/\sigma=1$, $\sigma_{sc}/\sigma>0$, $\kappa/\epsilon=0$; see Figure \ref{fig:fig1b}.
\item[OP] Overlapping-Polymer model. Here $b/\sigma<1$, $\sigma_{sc}/\sigma=0$, $\kappa/\epsilon=0$;  see Figure \ref{fig:fig1c}.
\item[OPSC] Overlapping-Polymer-with-Side-Chains model. Here $b/\sigma<1$, $\sigma_{sc}/\sigma>0$, $\kappa/\epsilon=0$; see Figure \ref{fig:fig1d}.
\end{itemize}

The difference between energetic and entropic stiffness was noted earlier \cite{Skrbic16b} and confirmed by a later study \cite{Werlich17}.  Here we explore the consequences of this difference, and the influence of adding the energetic stiffness to the OP and OPSC models.

We alert the reader about an important point. Our focus in this paper is on understanding the nature of the low temperature ($T^{*} \ll 1$)  behavior of the chain, a regime that will be referred to as the ’ ground state’. Under these conditions, the folding of the polymer is energy-dominated and the final ground state is the conformation with the lowest energy, which maximizes the number of favourable bonds, while not incurring excessive bond bending energy cost. However, sterics and other restrictions that were lumped under the name of entropic stiffness prevent the chain from adopting conformations that might yield an overall lower energy. This thinning of the allowed conformational space leads to interesting and novel ground states.


\section{Simulation details}
\label{sec:simulations}
\subsection{Monte Carlo method}
\label{subsec:MC}
The challenge of understanding the distinct roles of the three types of stiffnesses is best met by carrying out a detailed systematic study of the ground state phase diagrams using Wang-Landau microcanonical Monte Carlo simulations.
The zero-temperature phase diagram was calculated by means of microcanonical Wang-Landau Monte Carlo (MC) simulations \cite{Wang01} with no low energy cut-off. The density of states $g(E)$ that are visited along the simulation was iteratively built, by filling consecutive energy histograms. The acceptance probability was chosen to promote moves towards less populated energy states, thus providing for increasing flatness of energy histograms with the length of the simulation and leading the search towards lower and lower energy states. Each time a configuration of lower total energy was recorded, it was saved in the trajectory for further analysis. The set of MC moves included both local-type moves, such as single-sphere crankshaft, reptation and end-point moves, as well as non-local-type moves, such as pivot, bond-bridging and back-bite moves. For every state point, between 2 and 5 ground-state trajectories were monitored, consisting of between 10$^9$ and 3$\times$10$^9$ MC steps per bead. For more details see \cite{Skrbic19b}.
\subsection{Algorithm for knot detection}
\label{subsec:algorithm}
We used the method of Ref. \cite{Micheletti06},  based on the KNOTFIND algorithm, to confirm the existence and determine the type of knot. The basic idea \cite{Micheletti11} is to analyse the desired conformation by suitable ring closure followed by calculating the Alexander determinants. In order to ensure reliability, we carried out the standard practice of using two alternative closure schemes \cite{Tubiana11,Millett05}. The first is a minimally interfering closure scheme \cite{Tubiana11} followed by the computationally more demanding stochastic version \cite{Millett05}. The confirmation of non-trivial entanglement is provided by non-trivial Alexander determinants in a majority of 100 stochastic closures. 
\section{Numerical results}
\label{sec:results}
\subsection{Ground state of the WLC,OP, and PSC models: Knotted phases}
\label{subsec:ground}
\begin{figure}[H]
  \centering
  \captionsetup{justification=raggedright,width=\linewidth}
  \begin{subfigure}{7cm}
    \includegraphics[width=\linewidth]{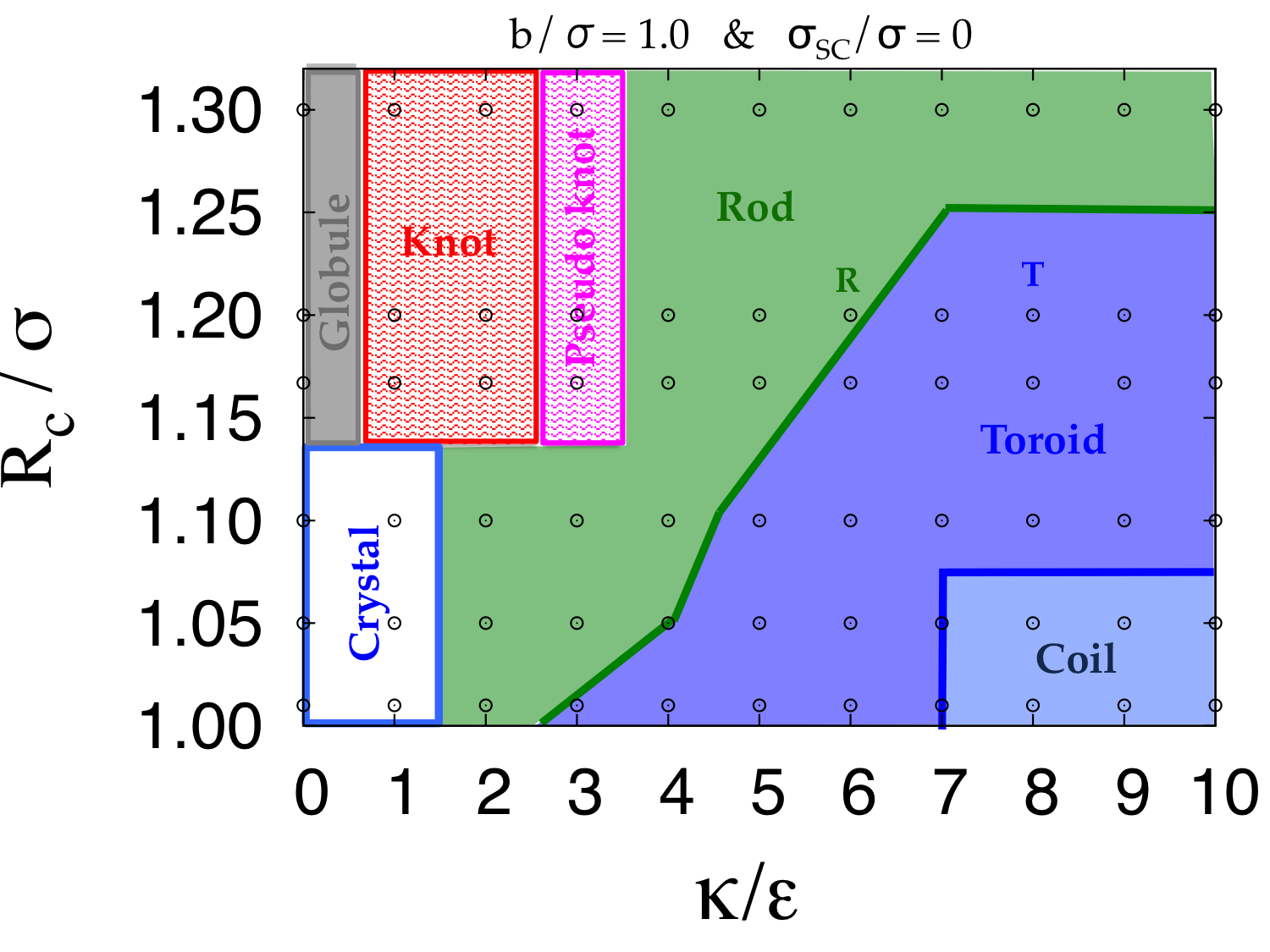}
    \caption{}\label{fig:fig2a}
  \end{subfigure}
    \begin{subfigure}{7cm}
    \includegraphics[width=\linewidth]{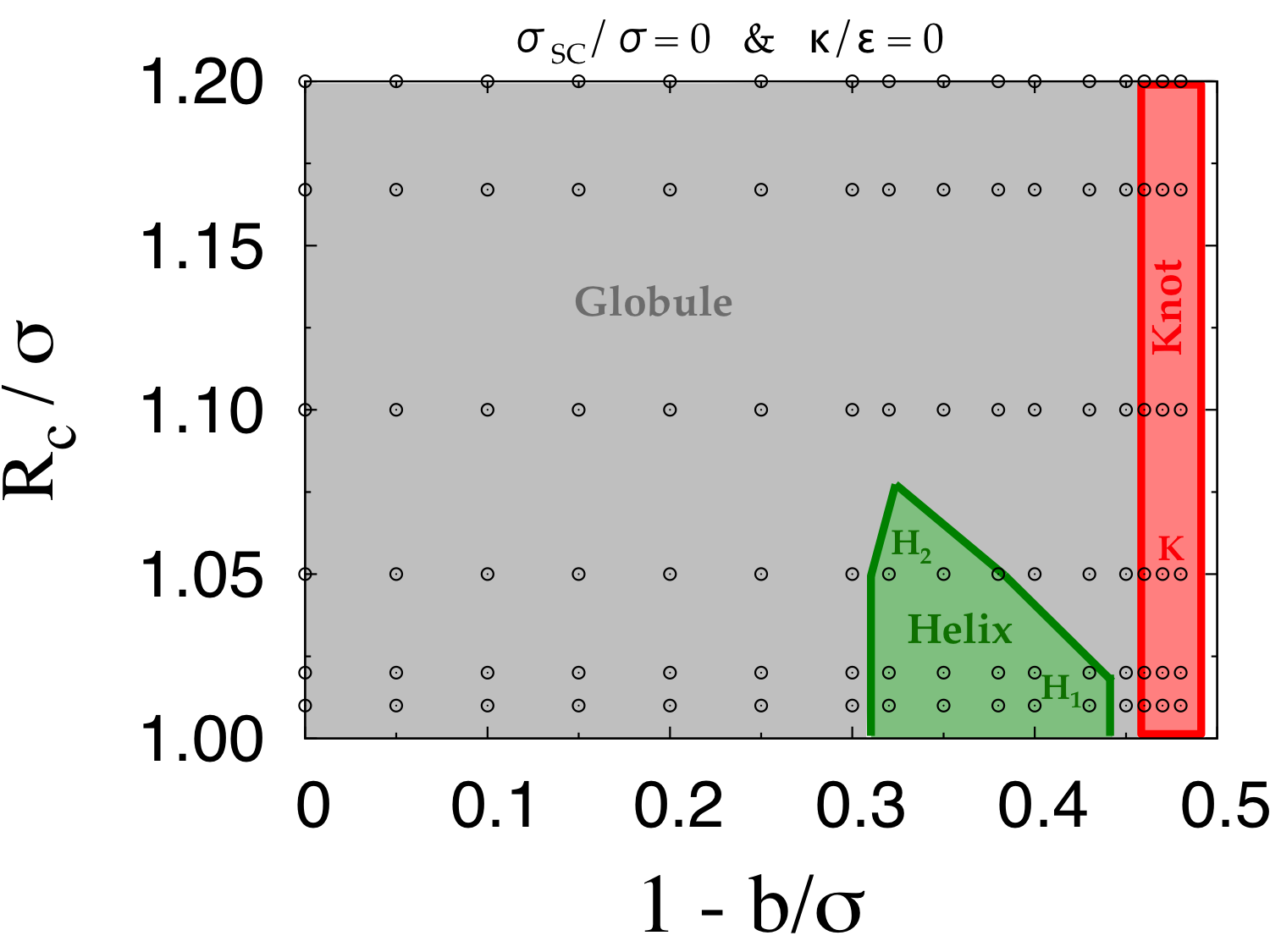}
    \caption{}\label{fig:fig2b}
  \end{subfigure}\\
  \begin{subfigure}{7.0cm}
    \includegraphics[width=\linewidth]{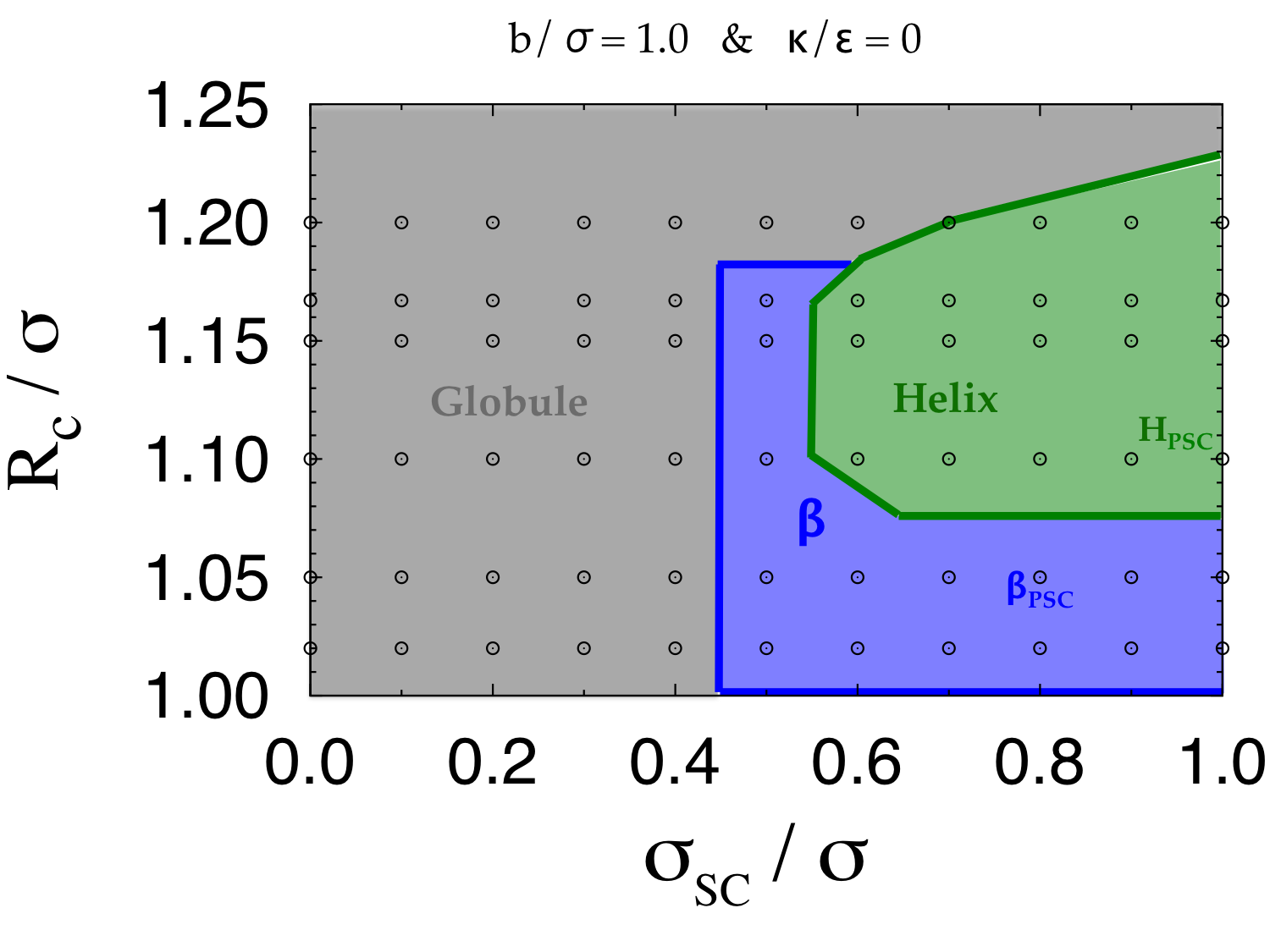}
    \caption{}\label{fig:fig2c}
  \end{subfigure}
    \begin{subfigure}{7.0cm}
    \includegraphics[width=\linewidth]{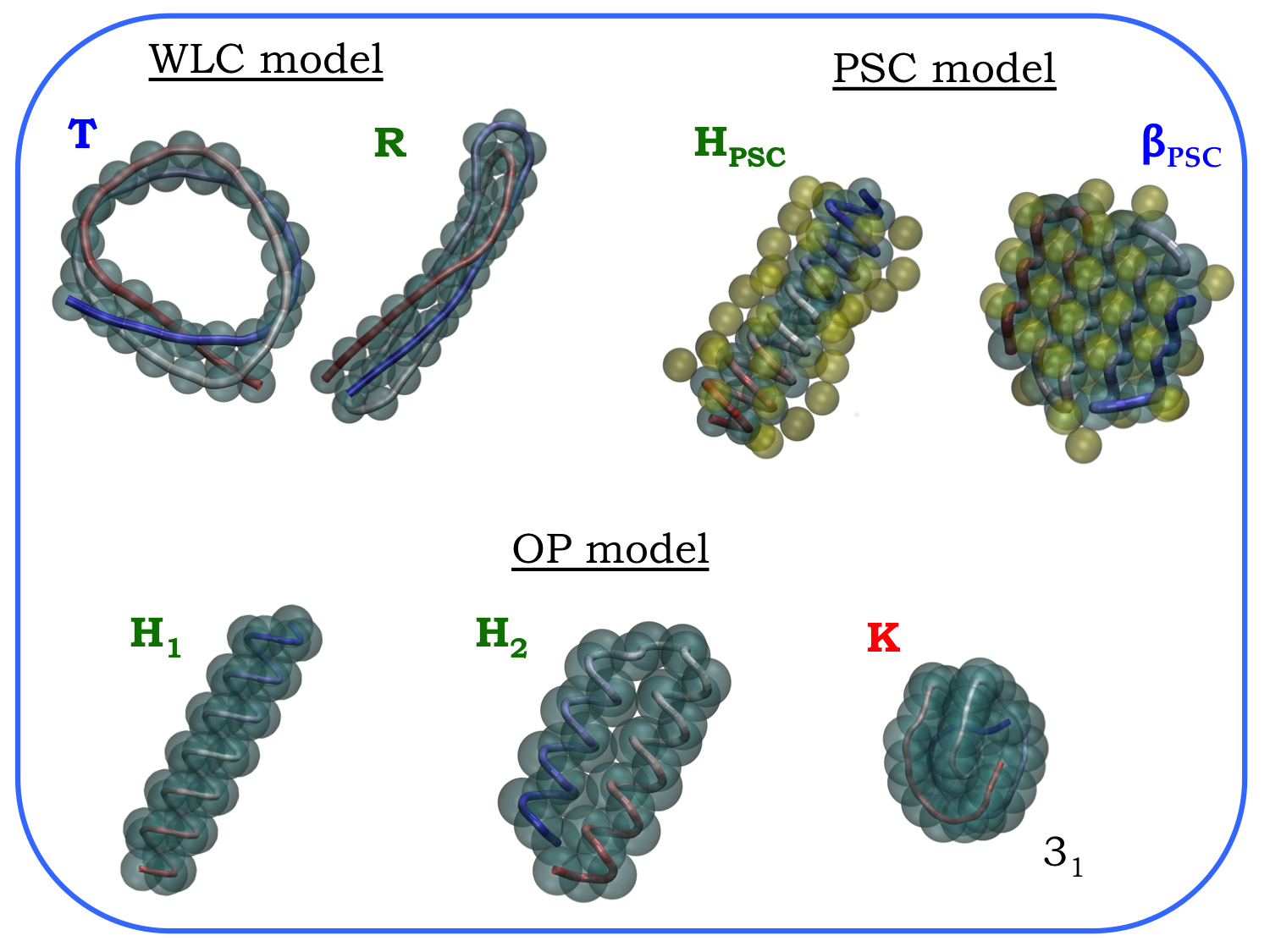}
    \caption{}\label{fig:fig2d}
  \end{subfigure}
    \caption{Ground state phase diagram of the (a) WLC model ($b/\sigma=1$) in the plane $R_c/\sigma$ versus $\kappa/\epsilon$; (b) OP model in the plane $R_c/\sigma$ versus $1-b/\sigma$; (c) PSC model in the plane $R_c/\sigma$ versus $\sigma_{sc}/\sigma$. (d) Snapshots of representative ground state conformations in the phase diagrams (a), (b) and (c) are shown.
  \label{fig:fig2}}
\end{figure}
Upon cooling down from a high temperature phase, a chain will tend to achieve a compact conformation (in order to maximally avail of the attractive interaction), whose nature depends upon the stiffness being considered. In the case of tangent beads ($b/\sigma=1$) there is only one parameter in the model ($R_c/\sigma$, the range of attractive interaction measured in units of the main chain sphere diameter) and a globular phase is achieved below a well-defined temperature (the $T_{\theta}$ temperature) whose value depends upon $R_c/\sigma$, and vanishes when $R_c/\sigma<1$.  In the case of tangent beads, there is a globular ground state phase when $R_c/\sigma>1$ and a coil phase when $R_c/\sigma<1$, and this is well-described by a Flory mean field theory \cite{Bhattacharjee13}.

The ground state of the WLC model exhibits structures, in addition to the compact (globule) and the swollen (coil) states, with characteristic conformations including toroids and rod-like folds \cite{Hoang14}. The configuration of parallel rods increases the number of energetically favorable contacts with a severe energetic stiffness penalty to accommodate the sharp backward bends facilitating the placement of parallel rods. For large stiffness, the backward bends are too expensive and the stiffness energy penalty is no longer localized but shared across contacts in the toroidal conformation.  The temperature phase diagram of the WLC model of a single chain subject to self-attraction was recently investigated by two different groups \cite{Seaton13,Marenz16}. The studies of the latter \cite{Marenz16} found knots, while the former \cite{Seaton13} did not. This difference was ascribed to the slightly different shapes of the attractive potentials \cite{Marenz16}.  Figure \ref{fig:fig2a} shows the corresponding ground state phase diagram that is fully consistent with the low temperature limit of Refs. \cite{Seaton13,Marenz16}. The key lesson learned is that knot formation was unfavorable when the stiffness was not large enough and when the range of attraction was not long enough. 

It is useful to distinguish between knots and pseudo-knots in the phase diagrams. A knot is one in which, loosely speaking, if one pulls the ends of the chain, the knot tightens and does not disappear. In contrast, a pseudo knot is one in which there is a knot but it can be made to vanish by pulling. The simplest mechanism for a pseudo-knot to form is to create a large enough toroidal hole within which a hairpin-like segment can penetrate. A smaller sized hole can only accommodate a single rod, which would then constitute a regular knot. 

The phase diagrams for the OP and the PSC models (Figures \ref{fig:fig2b} and \ref{fig:fig2c}) are distinct from that of the WLC model in Figure \ref{fig:fig2a} in several  ways. First, there are ground state phases with helical and $\beta$-structures in the models with entropic stiffness. These signatures of biomolecular phases arise from excluded volume effects and a thinning of phase space reminiscent of the pioneering work of Ramachandran and Sasisekharan \cite{Ramachandran68,Rose19b}. The OP model displays two different type of helices (\textbf{H$_{1}$} and  \textbf{H$_{2}$} (see representative snapshots in Figure \ref{fig:fig2d}), whereas the PSC model also shows a $\beta$-stranded phase for small values of $R_c/\sigma$ and larger values of side sphere sizes. Real protein helices are characterized by a pitch to radius ratio of $P/R \approx 2.4$. The helices in the OP and PSC models in general do not have these geometries, with the exception of the \textbf{H$_{2}$} helices in OP. The OP model exhibits a knotted phase in the limit of large entropic stiffness arising from a large overlap in accord with recent work \cite{Werlich17} -- the ground state was found to be a trefoil knot for sufficiently large interpenetration of adjoining spheres. We find that the knotted phase in the WLC model is very different from that in the OP model (we highlight this difference by employing different kinds of shading in the phase diagrams). 

The knotted phase in the WLC model is characterized by many low-lying states that are practically degenerate (within $5\%$ of the ground state energy). There is rich diversity in the flavor of knots and we have seen complicated topologies of knots including  $\textrm{4}_{\textrm{1}}$, $\textrm{5}_{\textrm{1}}$, $\textrm{5}_{\textrm{2}}$, $\textrm{6}_{\textrm{1}}$, $\textrm{6}_{\textrm{2}}$, $\textrm{6}_{\textrm{3}}$, $\textrm{7}_{\textrm{1}}$, $\textrm{7}_{\textrm{2}}$, $\textrm{8}_{\textrm{19}}$, $\textrm{8}_{\textrm{20}}$ and $\textrm{10}_{\textrm{124}}$ (see Fig.\ref{fig:fig5}).  The picture that emerges is that, on increasing the energetic bending stiffness,  random knots are formed in the vicinity of the globule phase along the route to the rod phase.  The knotted configurations arise as a balance between maximizing the number of contacts while not paying excessive cost associated with the energetic stiffness. Also, because the knots are formed randomly, there is no single robust ground state knot flavor as one might expect in a functionally vital protein phase. The energy landscape has many low lying local minima and is not conducive to reproducible folding. In contrast, $\textrm{3}_{\textrm{1}}$ knots are promoted almost exclusively by entropic stiffness and the ground state is much better separated in energy from the higher-energy states in accord with the observation \cite{Micheletti11,Chwastyk15,Alexander17} that more than 90\% of protein knots are $\textrm{3}_{\textrm{1}}$ in character.

\subsection{The WLC-OP model and its ground state}
\label{subsec:wlc-op}
\begin{figure}[htpb]
  \centering
  \captionsetup{justification=raggedright,width=\linewidth}
  \begin{subfigure}{8cm}
    \includegraphics[width=.8\linewidth]{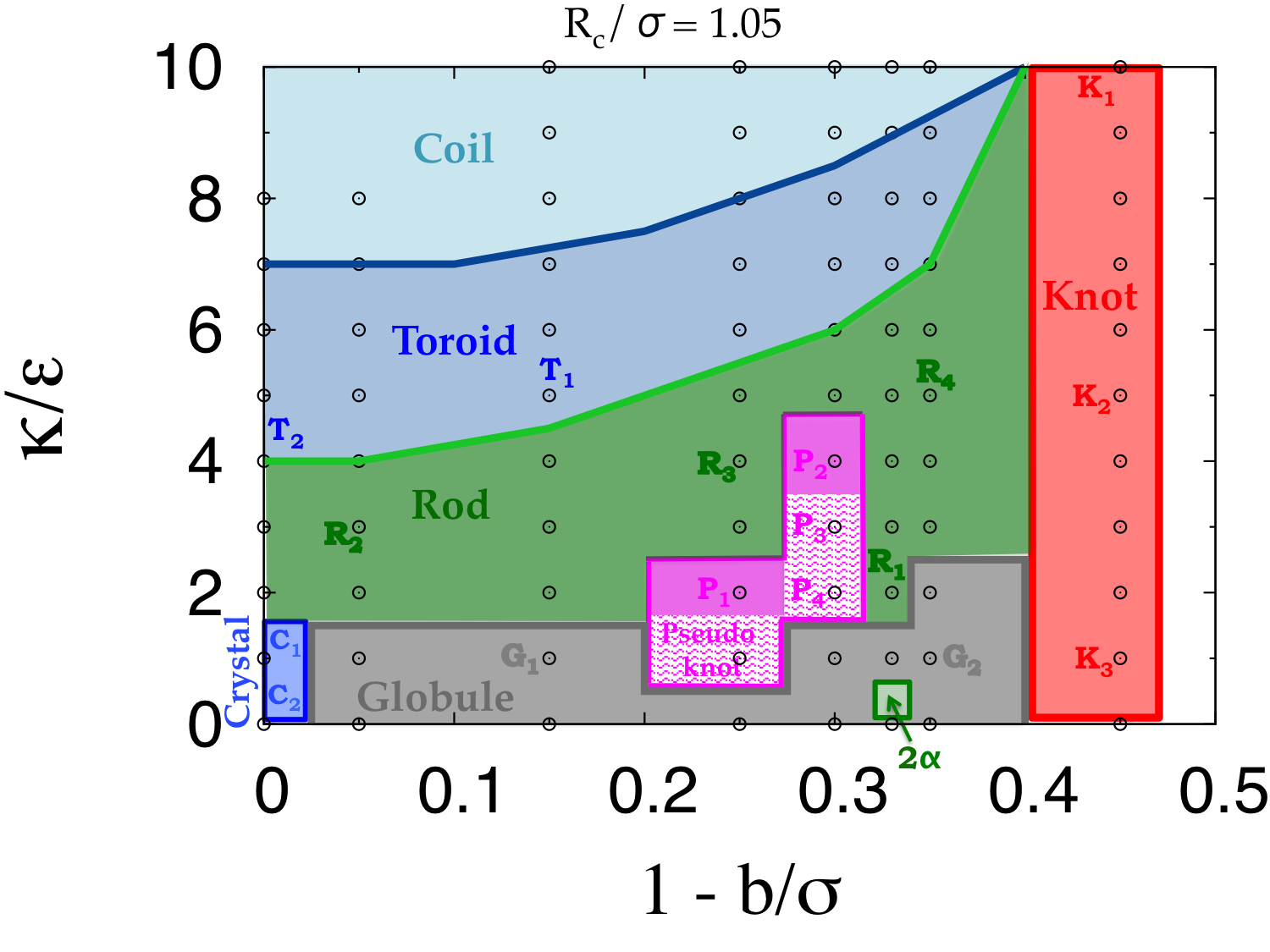}
    \caption{}\label{fig:fig3a}
  \end{subfigure}
  \begin{subfigure}{8cm}
    \includegraphics[width=.8\linewidth]{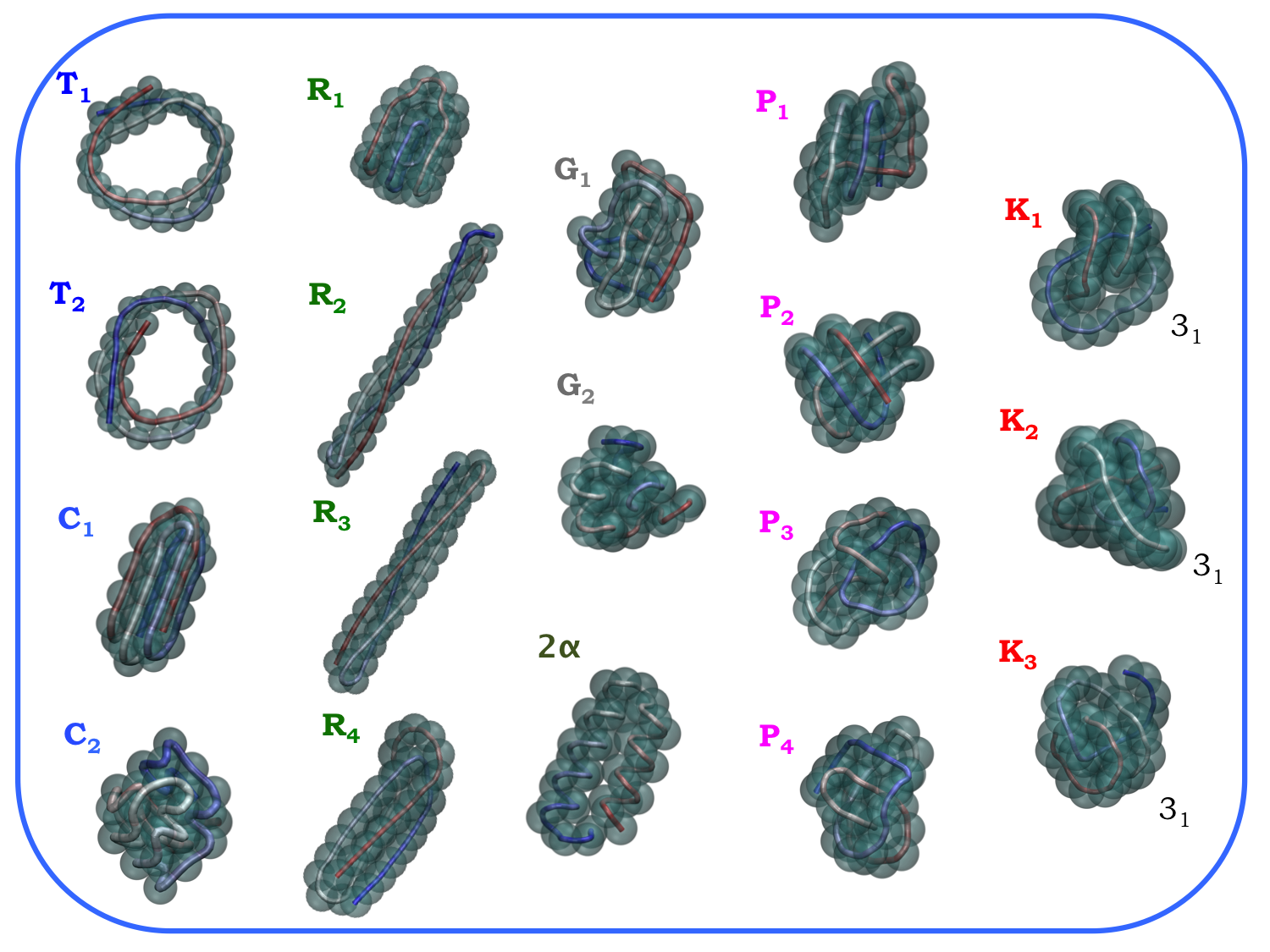}
    \caption{}\label{fig:fig3b}
  \end{subfigure}\\
  \begin{subfigure}{8cm}
    \includegraphics[width=.8\linewidth]{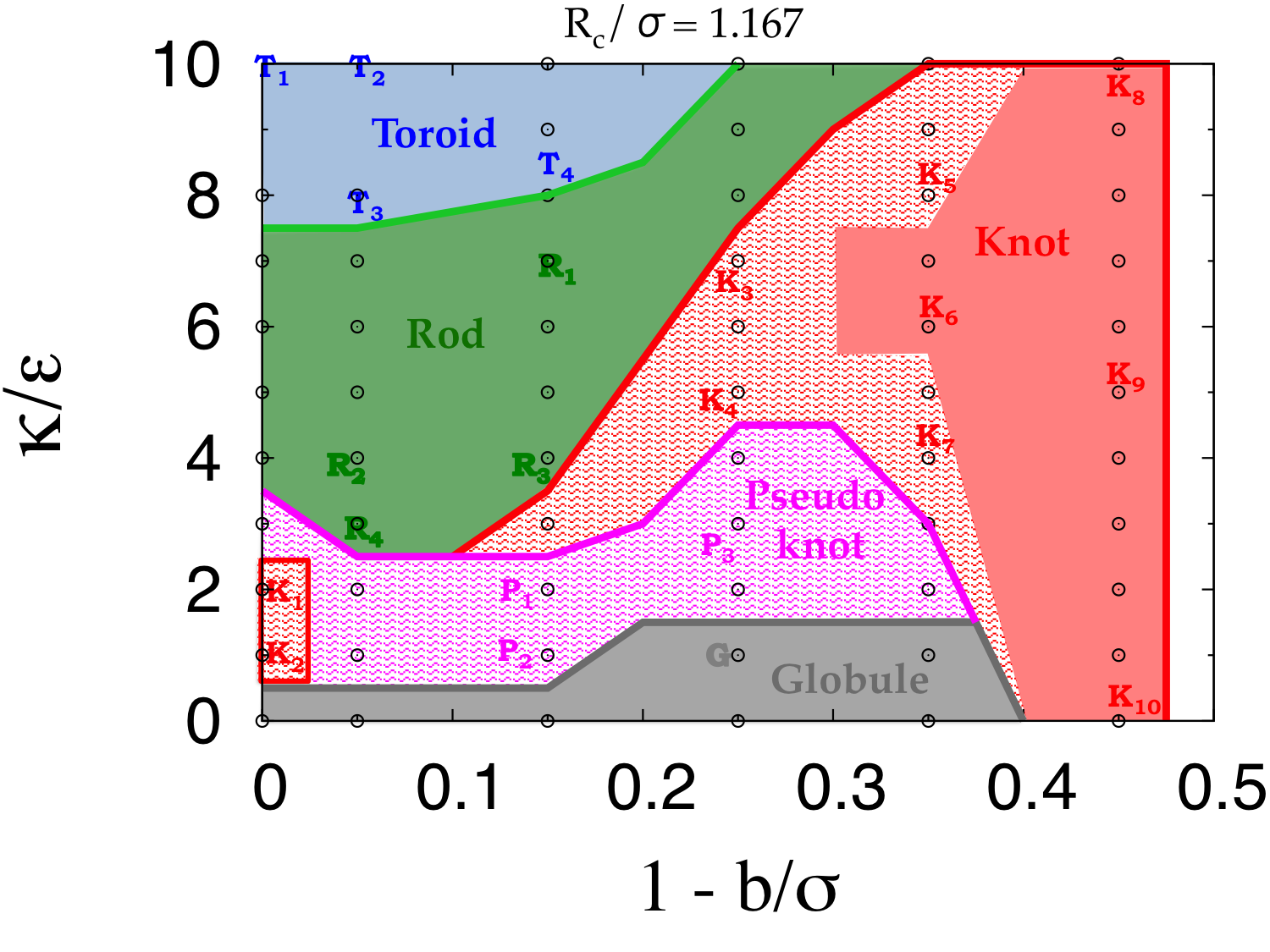}
    \caption{}\label{fig:fig3c}
  \end{subfigure}
    \begin{subfigure}{8cm}
    \includegraphics[width=.8\linewidth]{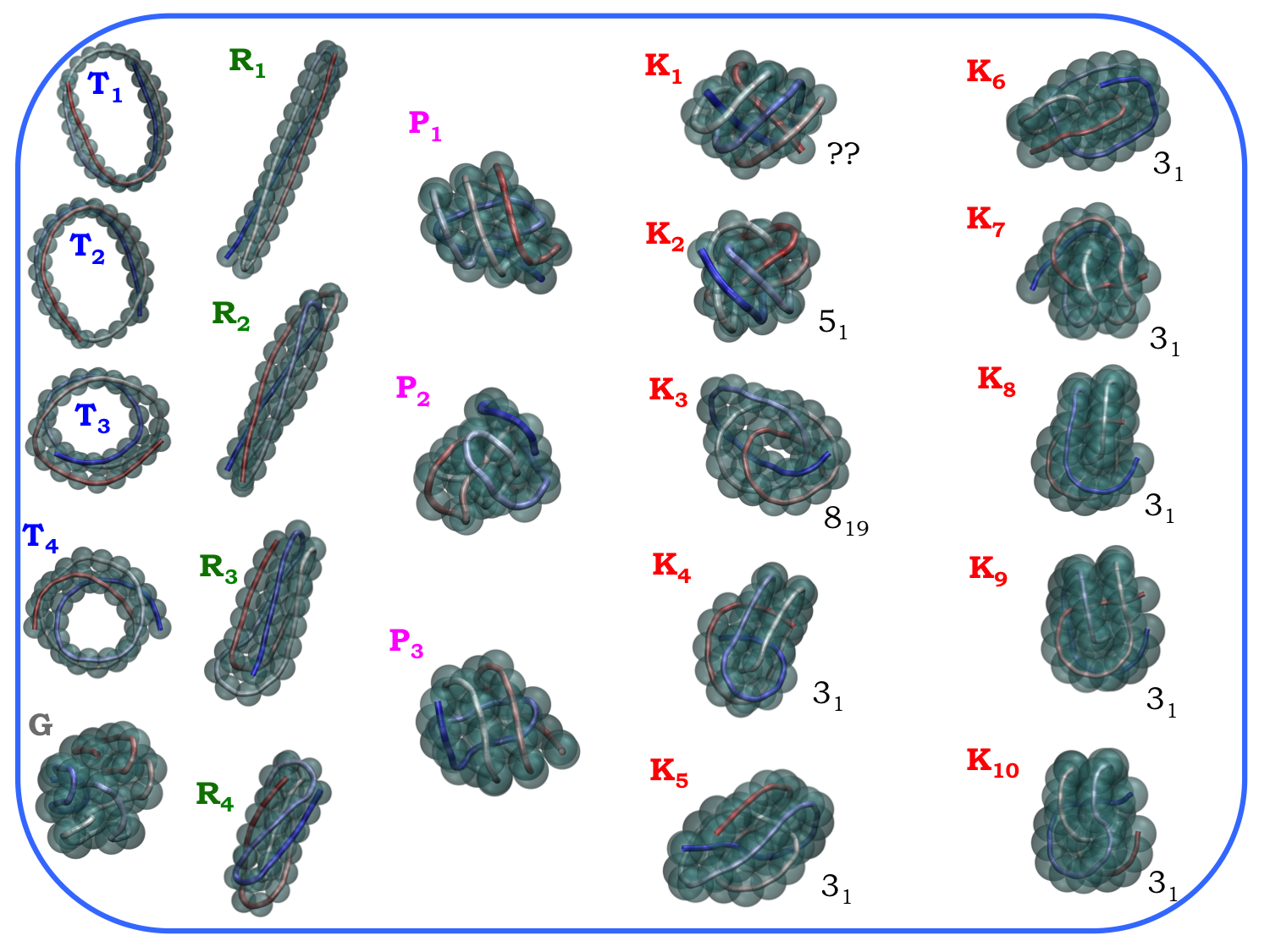}
    \caption{}\label{fig:fig3d}
    \end{subfigure} \\
      \begin{subfigure}{8cm}
    \includegraphics[width=.8\linewidth]{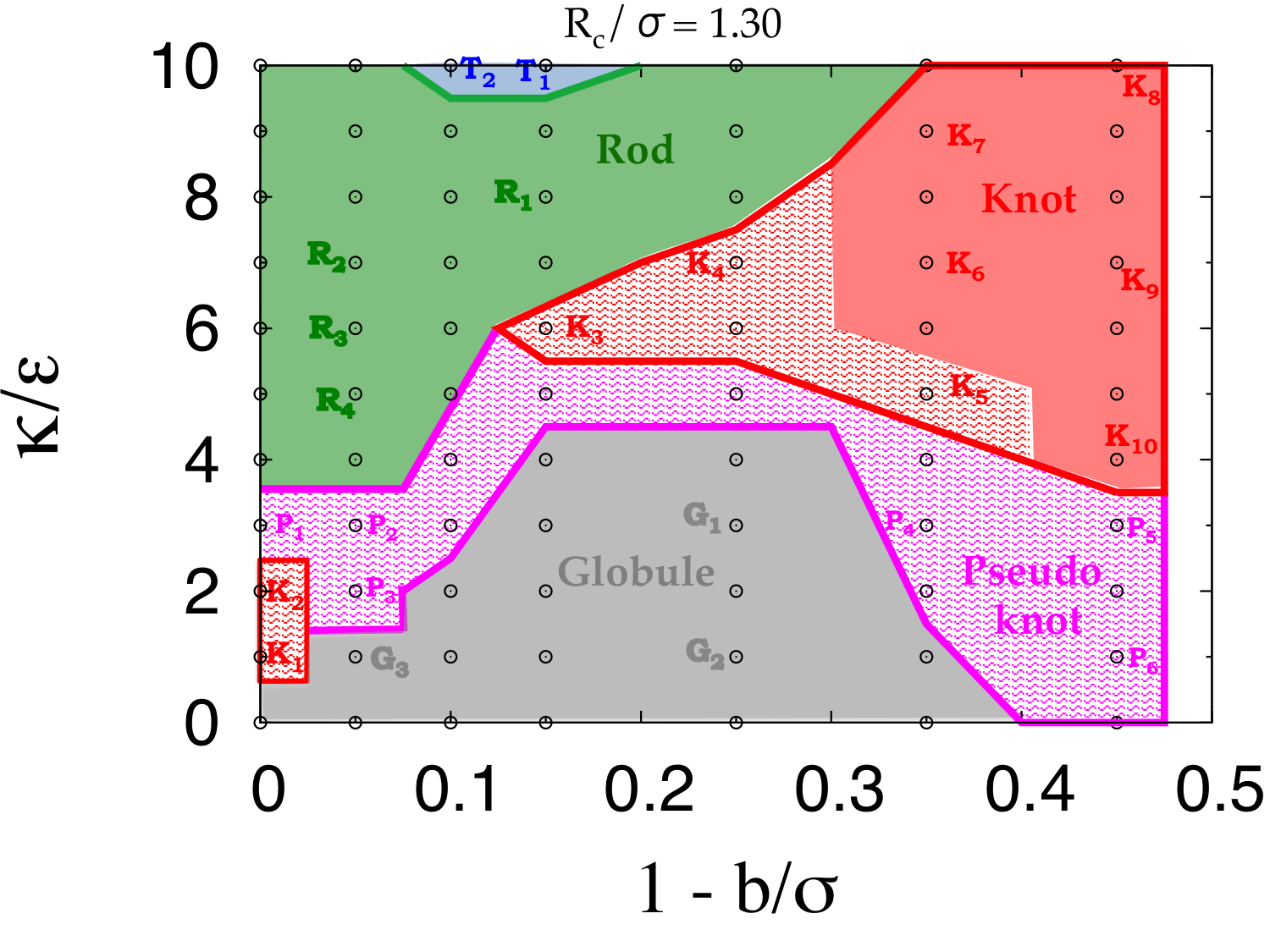}
    \caption{}\label{fig:fig3e}
      \end{subfigure}
        \begin{subfigure}{8cm}
    \includegraphics[width=.8\linewidth]{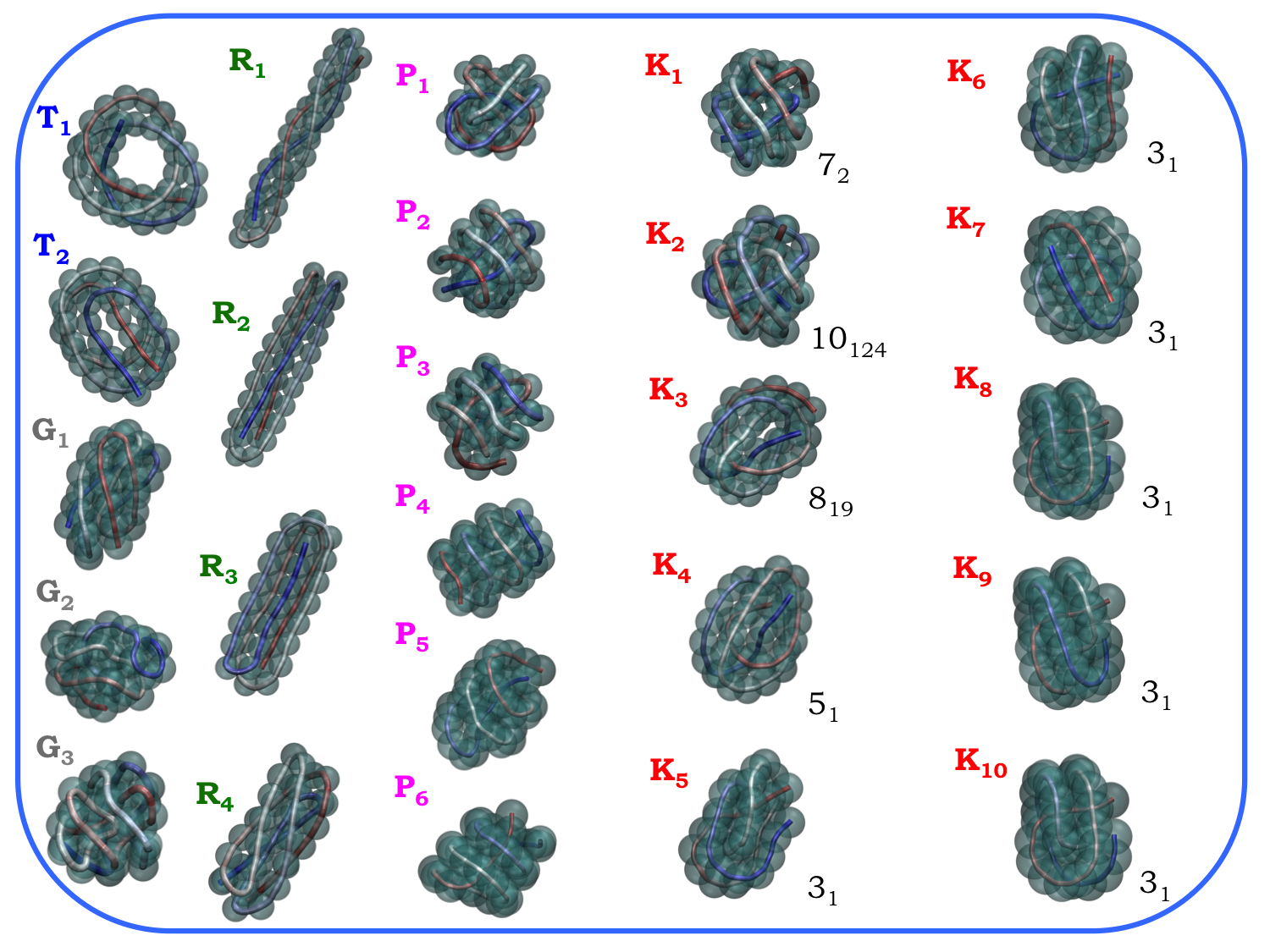}
    \caption{}\label{fig:fig3f}
  \end{subfigure}
  \caption{ Ground state phase diagrams of the combined WLC-OP (left panels) and representative snapshots of the corresponding conformations (right panel). Here the ground state phase diagrams are shown in the energy stiffness $\kappa$ --overlap fraction $1-b/\sigma$ plane for different ranges of attraction: (a)$R_c/\sigma=1.05$ ; (b) Representative snapshots for $R_c/\sigma=1.05$; (c) $R_c/\sigma=1.167$ ; (d)   Representative snapshots for $R_c/\sigma=1.167$; (e)  $R_c/\sigma=1.30$ ; (f)  Representative snapshots for  $R_c/\sigma=1.30$. The distinct phases are shown with different colors and the wave-like patterns indicate nearly degenerate ground states. The degeneracy manifests itself in two ways. First, in all parts of the phase diagram with a wave-like pattern, the energy landscape has many local minima with conformations having energies close to the ground state energy.  Such a landscape is not conducive to reproducible folding. The second type of degeneracy arises in the topologies of knotted low energy conformations. Here the trend is straightforward: close to the globular phase, one sees a variety of knot types, whereas there is fidelity in the flavour of knot when the entropic stiffness is large. This latter behaviour is akin to what is expected for protein native states. 
  \label{fig:fig3}}
\end{figure}

To gain further insight, we next consider the WLC-OP model which not only has the energy stiffness parameter of the WLC model but also allows consecutive spheres along the chain to partially overlap. This combined model has fewer degrees of freedom than the most general model incorporating all three types of stiffness (we will consider the latter briefly at the end of the paper). Figure \ref{fig:fig3} displays the ground state  phase diagrams in the plane bending rigidity $\kappa/\epsilon$, representing the worm-like-chain axis, versus interpenetration fraction $1-b/\sigma$, representing the overlapping-polymer axis for different ranges of attraction, $R_c/\sigma$, spanning very short range ($1.05$) to a value relevant to protein-like behavior ($1.16$) \cite{Skrbic19a,Skrbic19b} to an even larger value ($1.30$). Results for the individual overlapping-polymer and worm-like-chain models are obtained by setting $\kappa/\epsilon=0$ and $1-b/\sigma=0$.  In addition to the toroid and rod phases, knotted and pseudo-knotted phases are present in all cases. In the case of very short attraction range ($R_c/\sigma=1.05$), the knotted phase is confined within a long strip with high entropic stiffness ($1-b/\sigma \approx 0.5$) and different shapes of knots of the same flavour ($\textrm{3}_{\textrm{1}}$) are observed on increasing energetic stiffness (see Figure \ref{fig:fig3a} and Figure \ref{fig:fig6}). A pseudo-knotted phase is also present for $1-b/\sigma \approx 0.25$ and $\kappa/\epsilon$ in the interval 1-4 (see Figure \ref{fig:fig3a}). Note that the conventional globule phase pertaining to a tangent bead flexible chain ($1-b/\sigma \approx 0$ and $\kappa/\epsilon \approx 0$) is typically unstructured (except for crystalline order) and {\em not} knotted. As the range of attraction is increased   ($R_c/\sigma=1.167$, Figure \ref{fig:fig3c}), both the knotted and pseudo-knotted phases significantly expand along both axes with the emergence of additional knot types (Figure \ref{fig:fig3d} and Figure \ref{fig:fig7}).  At even larger range ($R_c/\sigma=1.30$, Figure \ref{fig:fig3e}) the toroid phase shrinks and the globular phase becomes dominant as one would expect on physical grounds. Interestingly, here we do observe knots and pseudo-knots in the flexible chain limit ($1-b/\sigma \approx 0$) and for small energetic stiffness ($\kappa/\epsilon \approx 1-3$).

Irrespective of the range of attraction, one important feature of the phase diagram of Figure \ref{fig:fig3} is related to the asymmetry in the promotion of knots from the two kinds (energetic -- related to $\kappa/\epsilon$, entropic -- related to $1-b/\sigma$) of stiffness. Essentially for any value of $\kappa/\epsilon$ (including the flexible limit $\kappa/\epsilon=0$), knot formation is achieved when the overlap $1-b/\sigma$ is maximal, demonstrating that the entropy stiffness \textit{always} tends to promote the formation of knots. In fact, even when knots are present at intermediate values of the overlap $1-b/\sigma \approx 0.3$, they tend to disappear upon increasing $\kappa/\epsilon$, a clear indication that energy stiffness tends to inhibit the formation of knots.

Our simulations have been carried out with great care using state-of-the-art computational techniques and paying particular attention to the onset of possible kinetic trapping. Hence, we are confident that the knotted conformations are indeed the true ground state.

\subsection{Energetic stiffness remove the degeneracy of the elixir phase}
\label{subsec:energetic}
\begin{figure}[htpb]
  \centering
    \includegraphics[width=.5\linewidth]{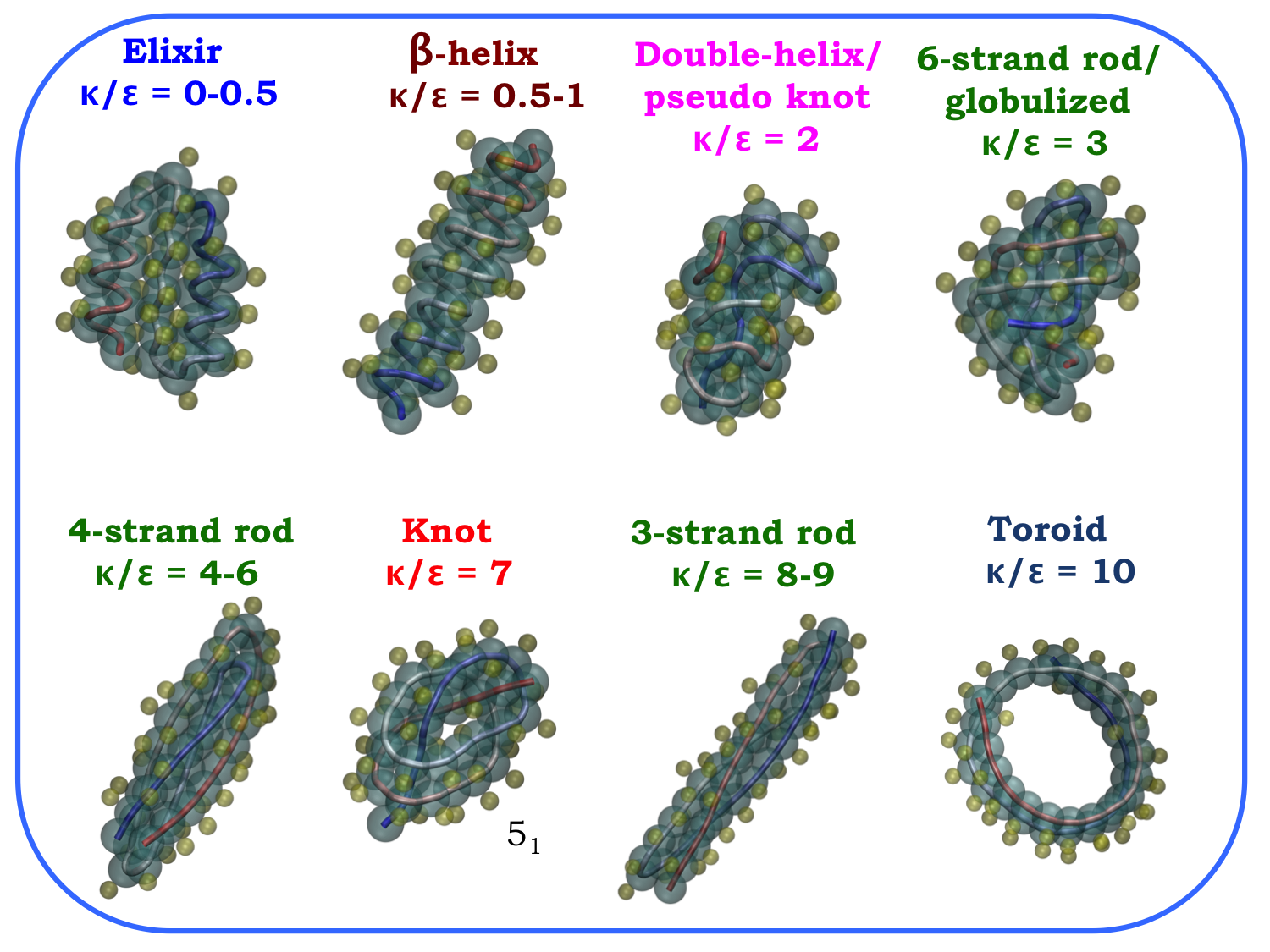}
    \caption{Conformational changes of a structure belonging to the elixir phase on increasing the energetic stiffness $\kappa/\epsilon$. Here $N=40$, $b/\sigma=0.75$, $R_c/\sigma=1.167$, and $\sigma_{sc}/\sigma =0.5$.
  \label{fig:fig4}}
\end{figure}
While both the OP and PSC models, whose phase diagrams have been displayed in Figures \ref{fig:fig2b} and \ref{fig:fig2c} respectively, are interesting in their own right, a combination of the two (the overlapping-polymer-with-side-chains OPSC model) has been shown to provide a very rich and instructive phase diagram \cite{Skrbic19a,Skrbic19b}, including characteristic folds \cite{Levitt76} appearing in real proteins. Of particular interest, is the so-called ``elixir phase'' that is located at the intersection of three different phases, the helix, the $\beta$, and the globular phase.  We refer the reader to Refs. \cite{Skrbic19a,Skrbic19b} for a detailed discussion of this phase diagram, and how the superstructures found in the elixir phase display remarkable similarities with those found in real proteins.

We focus here on the stability and nature of the ground state conformation on incorporating  the energetic stiffness into the OPSC model. A key finding is that the elixir phase is stable to the addition of weak energetic stiffness. An important characteristic of the elixir phase is the approximate degeneracy of the number of contacts in conformations made up of helices and strands. This attribute persists up to a threshold value of $\kappa/\epsilon \approx 0.5$. Figure \ref{fig:fig4} (see also Figure \ref{fig:fig8d}) shows the changes in the nature of the ground state conformation starting from a combined $\alpha-\beta$ structure in the elixir phase and accounts for the absence of the elixir phase in studies of a chain molecule incorporating the energetic stiffness.   

%
\subsection{How do different stiffness promote the formation of knots?}
\label{subsec:competition}
\begin{figure}[htpb]
 \centering
 \captionsetup{justification=raggedright,width=\linewidth}
   \includegraphics[width=.5\linewidth]{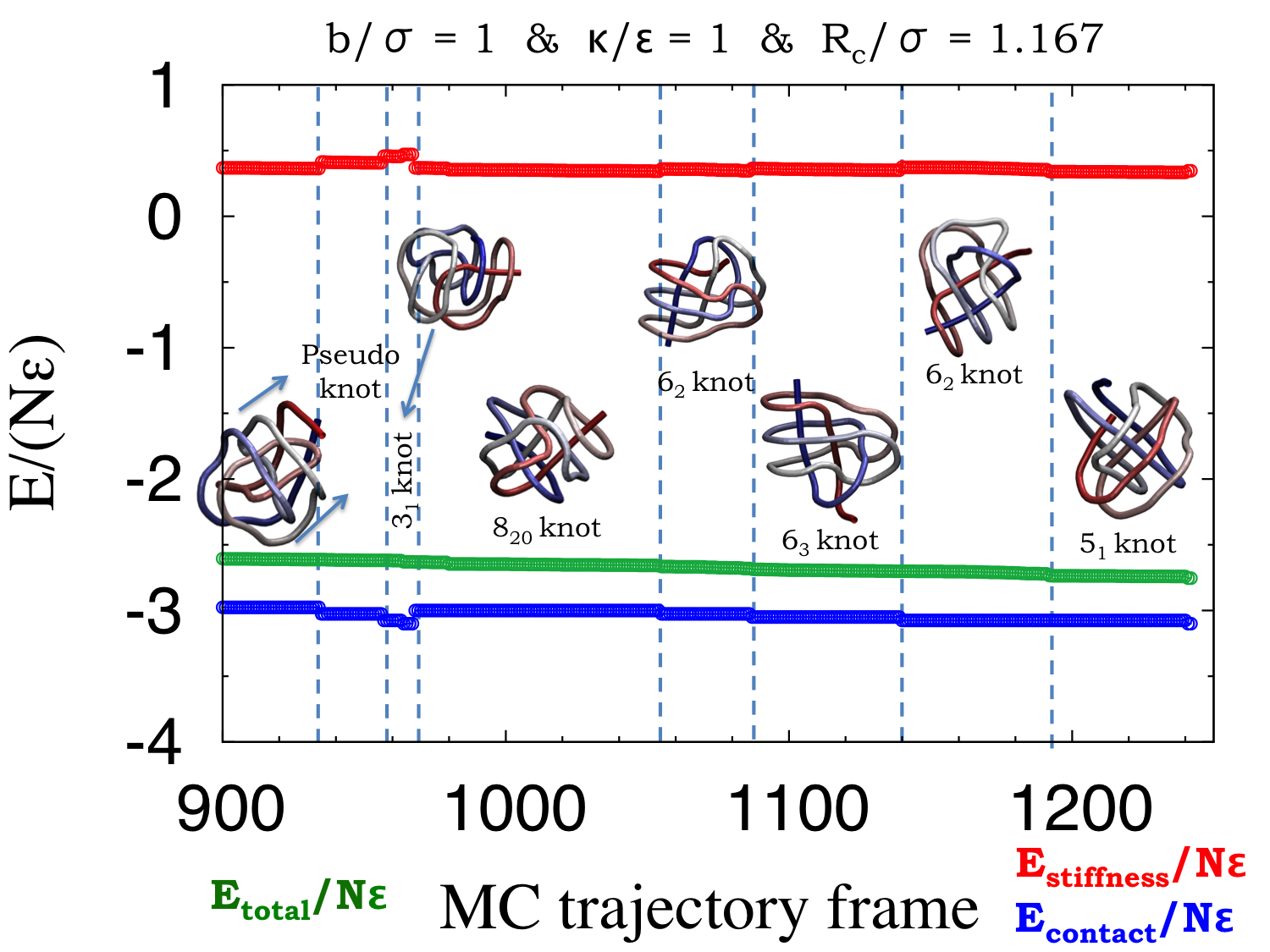}
	\caption{Energy per monomer (green line) of the WLC model during a single Wang-Landau trajectory (note that the x-axis does not simply represent a linear function of time, but success in finding a lower energy configuration which happens more infrequently as time progesses) of the ground state search for the case of $N=40$, $b/\sigma$ = 1, $\kappa$/$\epsilon$ = 1 and R$_c$/$\sigma$ = 1.167. This corresponds to state point $K_2$ in Figure \ref{fig:fig3c}. The red and blue lines correspond to the bending and contact energy per monomer, respectively. 
 \label{fig:fig5}}
\end{figure}
\begin{figure}[htpb]
 \centering
 \captionsetup{justification=raggedright,width=\linewidth}
	\includegraphics[width=.5\linewidth]{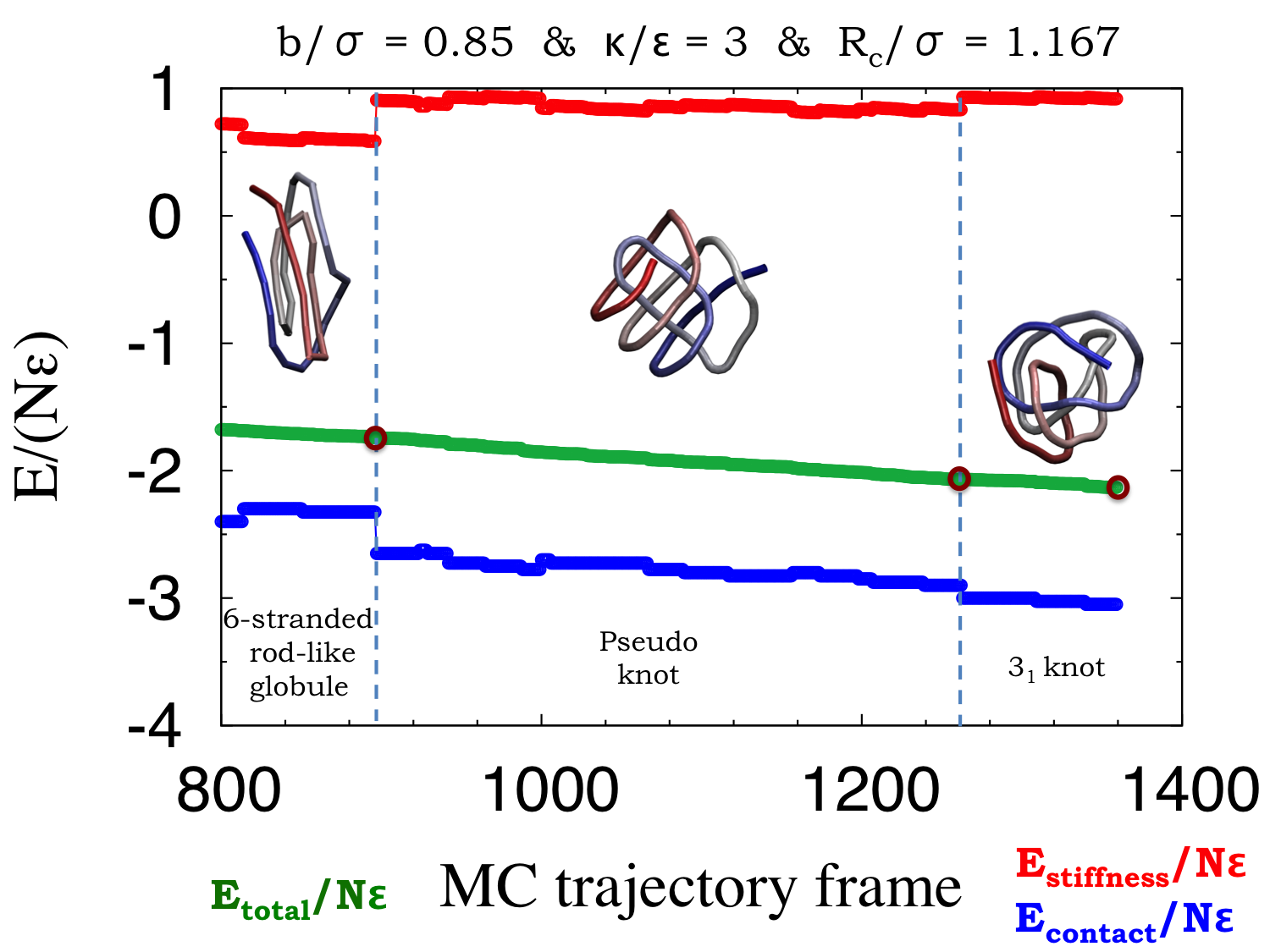}
	\caption{Energy per monomer of the WLC-OP model during a Monte Carlo trajectory  for the case $N=40$, $b/\sigma$ = 0.85, $\kappa$/$\epsilon$ = 3 and R$_c$/$\sigma$ = 1.167. As in Fig.\ref{fig:fig5} the x-axis does not simply represent a linear function of time, but success in finding a lower energy configuration which happens more infrequently as time progresses.
 \label{fig:fig6}}
\end{figure}
 We have seen that energetic and entropic stiffness have rather different roles in promoting the formation of knots in the ground state conformation of the chains. In order to get further insights, it proves instructive to disentangle the effect of the energetic penalty to be paid upon bending, and the energetic gain achieved by increasing the number of favourable contacts. We do this in Figure \ref{fig:fig5} for the WLC model, whose phase diagram is reported in Fig.\ref{fig:fig2a}.
   Here we report the total energy per monomer of the WLC model during a single Wang-Landau trajectory of the ground state search for the case of $N=40$, $b/\sigma$ = 1, $\kappa$/$\epsilon$ = 1 and R$_c$/$\sigma$ = 1.167. This corresponds to state point $K_2$ in Figure \ref{fig:fig3c}. We stress that here the x-axis does not simply represent a linear function of time, but success in finding a lower energy configuration which happens more infrequently as time progesses. Also depicted are the contact energy (number of favourable contact times $\epsilon$) per monomer, and the bending energy per monomer.
One observes conformations with distinct knot flavors, but with overall energies within 5\% of each other. The small kinks in the figure indicate the competing influences of maximizing the number of attractive contacts while keeping the stiffness energy in check, and entail the exploration of  distinct knot topologies.

\begin{figure}[htpb]
  \centering
  \captionsetup{justification=raggedright,width=\linewidth}
    \includegraphics[width=.5\linewidth]{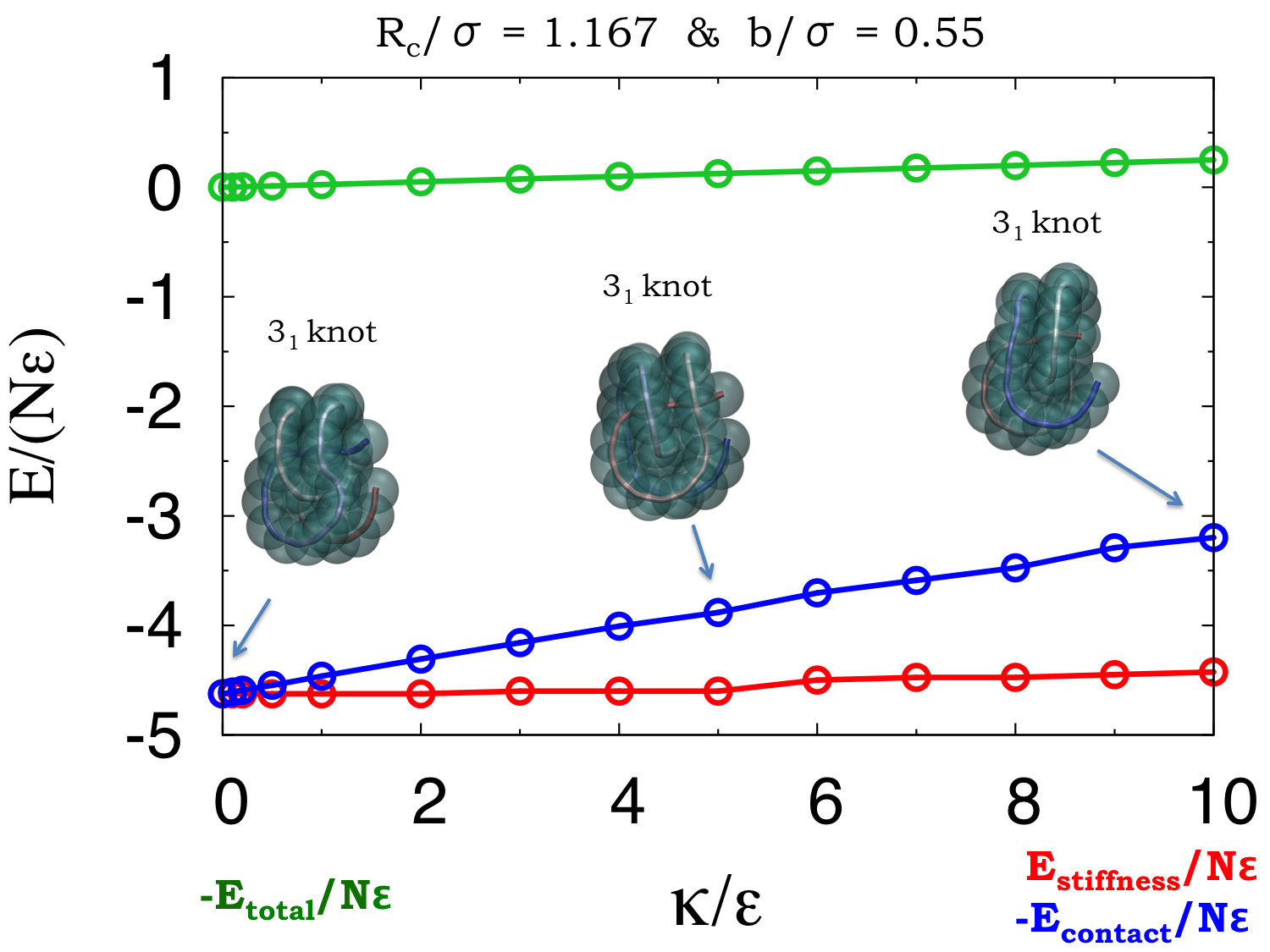}
	\caption{Ground state energy per monomer as a function of the stiffness parameter $\kappa/\epsilon$ of the WLC-OP model in the limit of very large  entropic stiffness. The green curve denote the ground state energy, the blue curve the attractive contact energy, and the red curve the stiffness penalty energy. 
  \label{fig:fig7}}
\end{figure}

A similar analysis is reported in Figure \ref{fig:fig6} in the case of the combined WLC-OP model. This corresponds to the knotted phase state point midway between the pseudo-knotted phase point $P_1$ and the rod-like phase point $R_3$ in Figure \ref{fig:fig3c}. For intermediate entropic stiffness, the number of distinct types of low-lying conformations is much smaller than the highly degenerate case of the tangent polymer (see Fig. \ref{fig:fig7}). Here we find a 3$_1$ knotted ground state that is very close in energy with pseudo-knotted conformations. The difference in the total energy between the lowest-energy pseudo-knot and the lowest total energy 3$_1$ configuration is less than 3\%. The energy of the next excited conformation (a 6-stranded rod-like globule) is 30\% higher than the 3$_1$ knotted ground state.

    The comparison between the results of Figure \ref{fig:fig5} (the WLC model with tangent beads) and Figure \ref{fig:fig6} (the WLC-OP model with
      non-tangent beads) underscores the role of the entropic stiffness in removing the degeneracy in the different knotted chain conformations having nearly identical energies. Further insights can be gained by monitoring the ground state energy per monomer as a function of the stiffness parameter of the WLC-OP model in the limit of very large  entropic stiffness. This is done in Figure \ref{fig:fig7}. The phase space trajectory being explored is along the straight line joining the points $K_{10}$, $K_9$ and $K_8$ in Figure \ref{fig:fig3c}. The three curves denote the ground state energy, the attractive contact energy, and the stiffness penalty energy. The latter has an approximate linear dependence on the stiffness parameter -- the nature of the ground state conformation (a 3$_1$ knotted structure) does not depend on the energy stiffness parameter. Note the convergence of the attractive contact energy and the stiffness penalty energy in the limit of tangent beads ($\kappa/\epsilon= 0$).

\begin{figure}[H]
 \centering
 \captionsetup{justification=raggedright,width=\linewidth}
 \begin{subfigure}{8cm}
   \includegraphics[width=.8\linewidth]{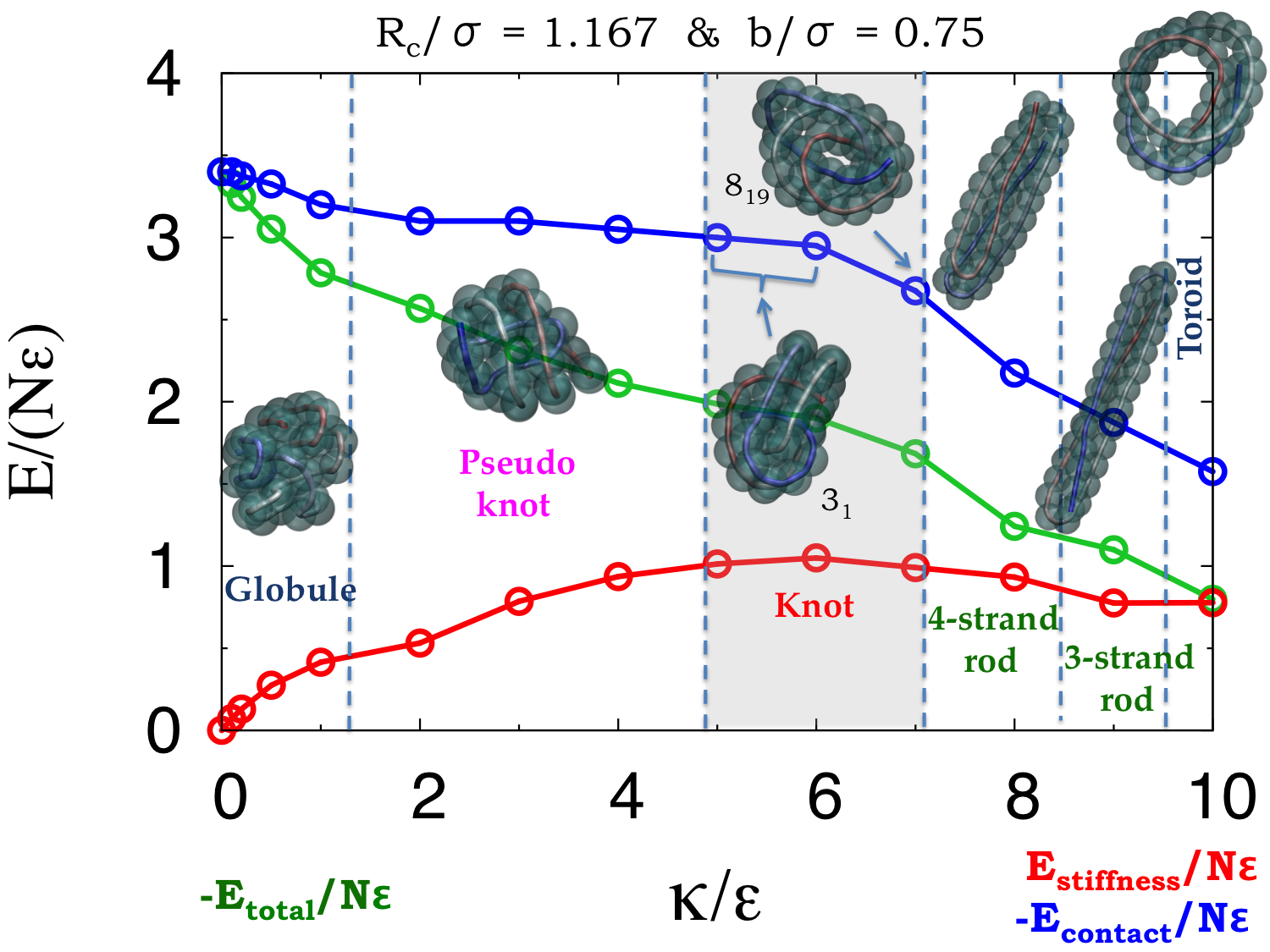}
   \caption{}\label{fig:fig8a}
 \end{subfigure}
 \begin{subfigure}{8cm}
   \includegraphics[width=.8\linewidth]{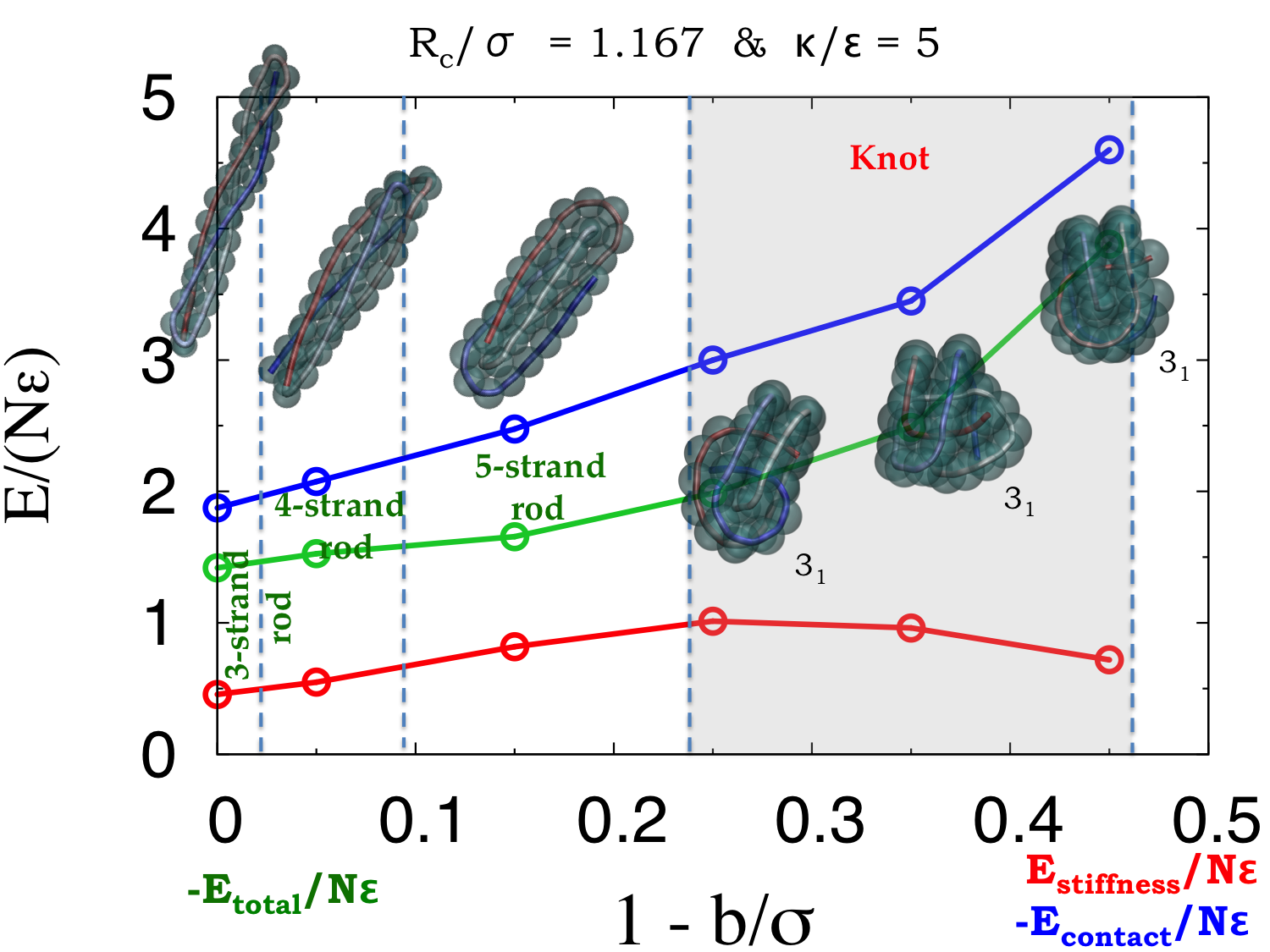}
   \caption{}\label{fig:fig8b}
 \end{subfigure}
 \begin{subfigure}{8cm}
   \includegraphics[width=.8\linewidth]{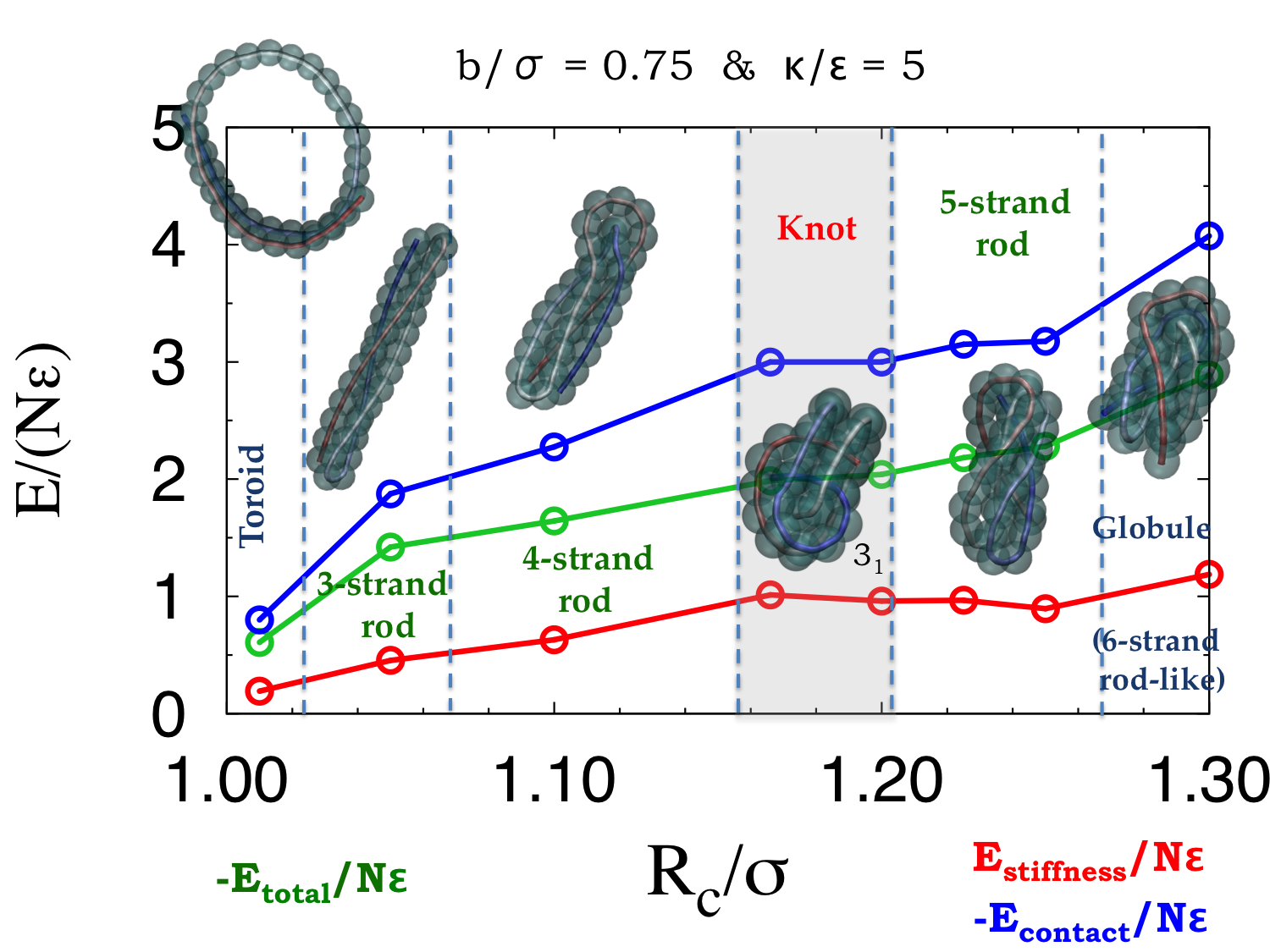}
   \caption{}\label{fig:fig8c}
 \end{subfigure}
 \begin{subfigure}{8cm}
   \includegraphics[width=.8\linewidth]{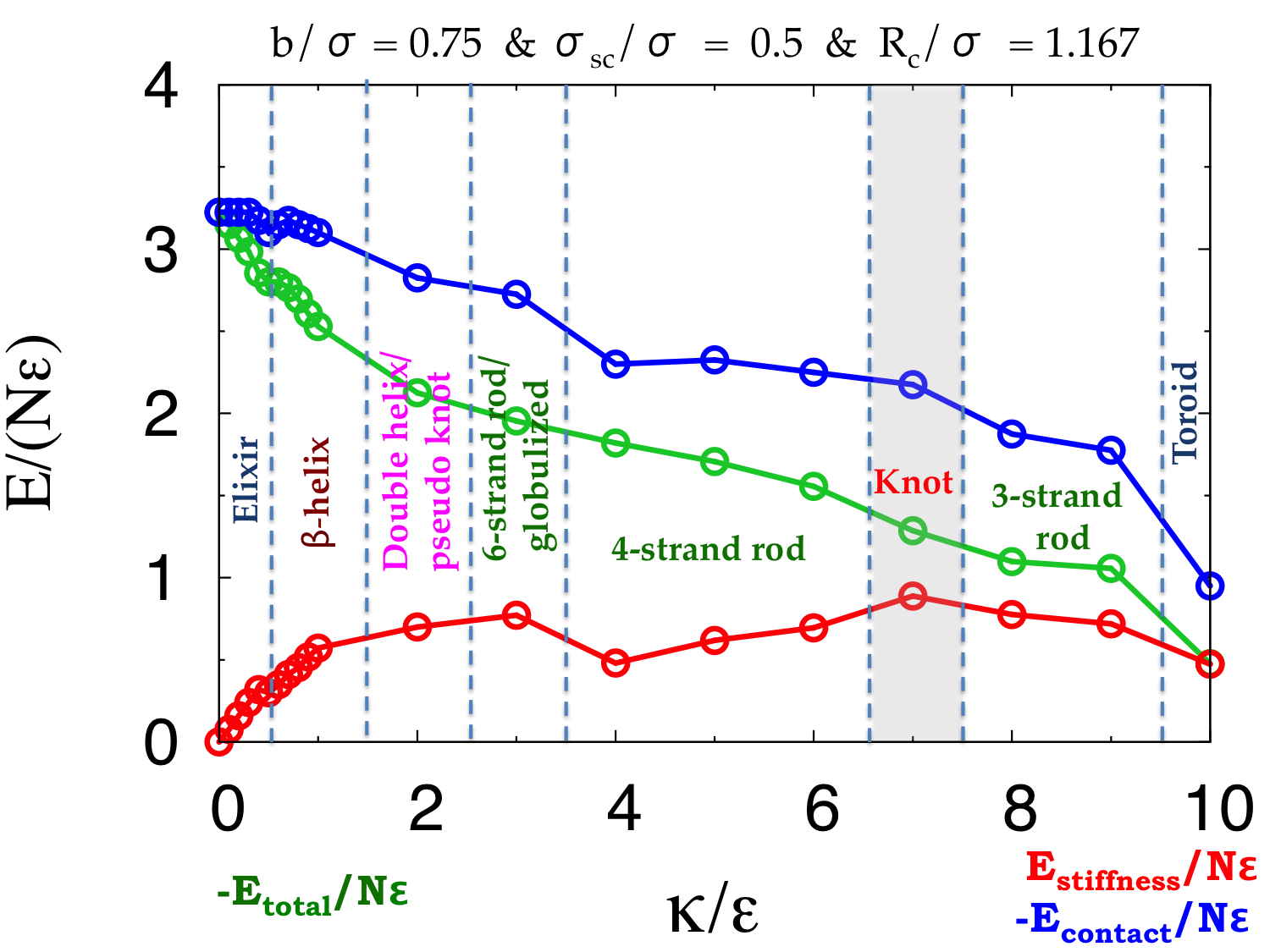}
	  \caption{} \label{fig:fig8d}
 \end{subfigure}
	\caption{(a) Energy per monomer $E/(N\epsilon)$ as a function of (a) the stiffness $\kappa/\epsilon$; (b) the fraction of overlap $1-b/\sigma$; and (c) the range of interaction $R_c/\sigma$ (c). Here $b/\sigma=0.75$ and $R_c/\sigma=1.16$ for (a); $R_c\sigma=1.16$ and $\kappa/\epsilon=5$ for (b);  and $b/\sigma=0.75$ and $\kappa/\epsilon=5$ for (c). (d) Energy per monomer $E/(N\epsilon)$ as a function of the stiffness $\kappa/\epsilon$ in the OPSC model. In all cases, we also plot the energy stiffness per monomer $E_{\text{stiffness}}/N$ (red curve) and the (negative) contact energy per monomer $E_{\text{contact}}/N$ (blue curve).
\label{fig:fig8}}
\end{figure}


We are now in the position of understanding the different effects of different parameters in the phase diagrams presented in Figure \ref{fig:fig2}. There are
  four parameters to deal with. The (reduced) bending energy $\kappa/\epsilon$; the fraction of overlap $1-b/\sigma$; the relative size of the side chain $\sigma_{sc}/\sigma$; and the relative range of square-well attraction $R_c/\sigma$. Figure \ref{fig:fig8} summarizes the results along these four parameters.
The first three panels refer to the WLC-OP model whereas the last panel is for the WLC-OPSC model. The first three panels show the energy as a function of the stiffness $\kappa/\epsilon$ (Figure \ref{fig:fig8a}); the range of attraction $R_c/\sigma$ (Figure \ref{fig:fig8b}; and the overlap of adjoining chain spheres $1-b/\sigma$ (Figure \ref{fig:fig8c}). The values of the two other parameters are shown at the top of each panel. Figure \ref{fig:fig8d} displays the energy per monomer as a function of the stiffness parameter $\kappa/\epsilon$  in the full WLC-OPSC model in the ``center'' of the elixir phase: $b/\sigma = 0.75$ and $\sigma_{SC}/\sigma = 0.5$. In all panels, the three  lines denote the total energy, the attractive contact energy, and the stiffness energy penalty. Interestingly, in all cases, the knotted configurations occur when stiffness energy penalty is at local maximum underscoring that the origin of knots arises from the interplay between the energetic cost of the stiffness and the energetic gain from the attractive contacts facilitated by knot formation.

\section{Different stiffnesses can yield the same persistence length $L_p$}
\label{sec:different}

The WLC model ascribes an elastic energy penalty to the local bending of a chain.  
Let $\widehat{\mathbf{T}}_i$ be the local tangent along the chain axis at the $i$-th bead, so that the bending angle $\theta_i$ between the
$i$-th and the $(i+1)$-th bead is given by Eq.\ref{supp1:eq3}.

The WLC model then considers the potential (\ref{supp1:eq4}) whose choice of the $\theta$ dependence ensures an elastic behavior for small $\theta$
as well as a flattening of this energy at larger angles and derives the persistence length -- the length below which the directions of different chain segments are correlated, as  
\begin{eqnarray}
  \label{supp3:eq1}
  \frac{L_P}{\sigma} &=& - \frac{1}{\log \left \langle \cos \theta \right \rangle}
\end{eqnarray}
Using the WLC potential (\ref{supp1:eq4}) we can compute the average $\langle \cos \theta \rangle$ as 
\begin{eqnarray}
  \label{supp3:eq2}
  \left \langle \cos \theta \right \rangle &=& \frac{\int_{-\pi}^{+\pi}~ d \theta~ \cos \theta e^{-\beta \kappa \left(1-\cos \theta \right)}}{\int_{-\pi}^{+\pi}~ d \theta ~e^{-\beta \kappa \left(1-\cos \theta \right)}}
\end{eqnarray}
With a few algebraic manipulations, this average can be reduced to
\begin{eqnarray}
  \label{supp3:eq3}
  \left \langle \cos \theta \right \rangle &=& \frac{I_{1}\left(\beta \kappa \right)}{I_{0}\left(\beta \kappa \right)}
\end{eqnarray}
where $I_1(z)$ and $I_0(z)$  are Bessel functions of integer order \cite{Abramowitz65}
\begin{eqnarray}
  \label{supp3:eq4}
  I_n\left(z\right) &=& \frac{1}{\pi} \int_{0}^{+\pi}~ d\theta\, \cos\left(n\theta\right) e^{z \theta}
\end{eqnarray}
For low temperatures $T^{*} \ll 1$, we can use the asymptotic expansion \cite{Abramowitz65}
\begin{eqnarray}
  \label{supp3:eq5}
   I_n\left(z\right) &=& \frac{e^z}{\sqrt{2 \pi z}} \left[1-\frac{\left(4n^2-1\right)}{8z} + {\cal O}\left(\frac{1}{z^2}\right) \right]
\end{eqnarray}
so that
\begin{eqnarray}
  \label{supp3:eq6}
  \frac{I_1\left(z\right)}{I_0\left(z\right)} &=& 1 - \frac{1}{2z} +  {\cal O}\left(\frac{1}{z^2}\right)
\end{eqnarray}
This leads to
\begin{eqnarray}
  \label{supp3:eq7}
 \left \langle \cos \theta \right \rangle &\approx& 1 -\frac{1}{2 \frac{k}{\epsilon}\frac{1}{T^{*}}}
\end{eqnarray}
From the above average one can then obtain the persistence length $L_p$ through Eq.(\ref{supp3:eq1}).

Following the suggestion given in Ref.\cite{Toan05}, we now show that the same calculation can be carried out for the case of entropic stiffness, taking the OP as an example. Here the
potential has the all-none form
\begin{equation}
\phi_{\text{OP}}\left(\theta\right)=\begin{cases}
+\infty\,,\quad \,\theta < \theta_{0} &\\
0, \qquad \theta > \theta_{0}&
\end{cases}
\label{supp3:eq8}
\end{equation}
where  $\theta_0$ is the maximum bending angle that clearly depends upon the ratio $b/\sigma$ as
\begin{eqnarray}
  \label{supp3:eq9}
  \cos \frac{\theta_{0}}{2} &=& \frac{\sigma}{2b}
\end{eqnarray}
so that, using Eq.(\ref{supp3:eq2}) we have 
\begin{eqnarray}
  \label{supp3:eq10}
  \left \langle \cos \theta \right \rangle &=&
  \frac{\int d\theta \cos \theta e^{-\beta  \phi\left(\theta\right)}}{\int d\theta  e^{-\beta  \phi\left(\theta\right)}} \\
  &=& \frac{\int_{\cos \theta_{0}}^{1} d\left(\cos \theta\right) \cos \theta}{\int_{\cos \theta_{0}}^{1} d\left(\cos \theta\right)}=
  \frac{\sigma^2}{4 b^2}
\end{eqnarray}
and hence
\begin{eqnarray}
  \label{supp3:eq11}
 \left \langle \cos \theta \right \rangle &=&  \frac{\sigma^2}{4 b^2}
\end{eqnarray}
Note that this result does not depend upon the temperature.
Upon comparing Eqs.(\ref{supp3:eq11}) and (\ref{supp3:eq7}), one then obtains the condition under which the two persistence lengths are the same  
\begin{eqnarray}
  \label{supp3:eq12}
  \frac{\kappa}{\epsilon} &=& \frac{T^{*}/2}{1-\frac{1}{4b^2/\sigma^2}}
\end{eqnarray}
that corresponds to a line in the ($\kappa/\epsilon$)-($1-b/\sigma$) plane. Note that this correspondence is strictly valid only in the low-temperature limit.

In Ref. \cite{Skrbic16b}, it was already noted that it was possible to modify the WLC stiffness potential (\ref{supp1:eq4}) to match Eq.(\ref{supp3:eq8}) that can be ascribed to the OP model, with the result that the persistence length numerically computed in the two cases turned out to be the same (the average is the thermal average).

The key point of this analysis is that the persistence length alone is insufficient to capture the nuances of different type of stiffnesses.

\section{Conclusions}
\label{sec:conclusions}
We have presented the results of detailed studies aimed at understanding the rich interplay between different kinds of stiffnesses in determining the ground state conformation of a chain molecule subject to self-attraction. Remarkably, the conformations range from an unstructured globule to almost parallel rods to a toroid and, even more surprisingly, assembled structures built of helices and strands resembling protein native state structures. The physical factors underlying the formation of these diverse conformations can be understood based on the nature of the stiffness. Equally interestingly, our study sheds light on the question of the origin and types of knotted structures even for relatively short chains. Taken together, these provide hints on the origin of knotted native state structures of proteins into which reproducible folding takes place. In the limit of maximal entropic stiffness, our studies point to the passive role played by the energetic stiffness, whereas for intermediate values of the entropic stiffness, there is a non-trivial role played by each type of stiffness (Figures \ref{fig:fig8a}, \ref{fig:fig8b}, \ref{fig:fig8c}). Studies of models combining different types of stiffnesses provide a holistic view on the nature of the constraints and the rich interplay between energetic costs (due to sharp bends), energetic gains of attractive contacts due to compaction, and entropic constraints thinning the phase space of conformations available to a chain molecule. While different kinds of stiffness can yield the same persistence length (as discussed in Section \ref{sec:different}), they can yet have distinct effects on the nature of the ground state conformation based on their physical origin. Entropy stiffness promotes the formation of knots in the ground state to increase the number of favorable attractive contacts while the energetic cost of sharp bends discourages knotting in the ground state structurs. We have demonstrated that the incorporation of distinct types of stiffness allows one to bridge conventional polymeric structures and biomolecular structures underscoring their unity.

\acknowledgments{We are indebted to Michele Cascella, Ivan Coluzza, Brian Matthews, Flavio Romano, George Rose, Francesco Sciortino, Luca Tubiana and especially Pete von Hippel, Trinh Hoang and Amos Maritan for invaluable discussions and/or collaboration.
We are indebted to Cristian Micheletti for sharing the software for knots detection and for much helpful advice.
  The use of the high performance computer Talapas  at the University of Oregon and the SCSCF multiprocessor cluster at  the Universit\`{a} Ca' Foscari Venezia are gratefully acknowledged. The work was supported by MIUR PRIN-COFIN2017 \textit{Soft Adaptive Networks} grant 2017Z55KCW (A.G.), and a G-1-00005 Fulbright and University of Oregon Research Scholarship (T.\v{S}.).}



\end{document}